\documentclass[11pt]{article}

\usepackage[margin=1in]{geometry}
\usepackage[T1]{fontenc}
\usepackage{lmodern}
\usepackage{xcolor}
\usepackage{placeins}
\usepackage[round,authoryear]{natbib}
\usepackage[colorlinks=true,linkcolor=blue,citecolor=blue,urlcolor=blue]{hyperref}
\usepackage{setspace}
\usepackage{amsfonts}       
\usepackage{nicefrac}       
\usepackage{microtype}      

\usepackage{epsf}
\usepackage{epsfig}
\usepackage{fancyhdr}
\usepackage{graphics}
\usepackage{graphicx}
\usepackage{mdframed}
\usepackage{tikz}
\usepackage{xparse}     
\usepackage{etoolbox} 

\usepackage{microtype}
\usepackage{graphicx}
\usepackage{booktabs} 
\usepackage{url}

\usepackage{mathrsfs}
\usepackage{dsfont}
\usepackage{amsmath}
\usepackage{amssymb,bbm}
\usepackage{caption}

\usepackage{wrapfig}
\usepackage{subcaption}

\usepackage{algorithm}
\usepackage{algpseudocode}

\long\def\comment#1{}
\definecolor{battleshipgrey}{rgb}{0.52, 0.52, 0.51}
\definecolor{darkgray}{rgb}{0.66, 0.66, 0.66}
\definecolor{darkgreen}{rgb}{0.0, 0.2, 0.13}
\definecolor{darkspringgreen}{rgb}{0.09, 0.45, 0.27}
\definecolor{dukeblue}{rgb}{0.0, 0.0, 0.61}
\definecolor{olivedrab7}{rgb}{0.24, 0.2, 0.12}
\definecolor{darkblue}{rgb}{0.0, 0.0, 0.55}
\definecolor{darkscarlet}{rgb}{0.34, 0.01, 0.1}
\definecolor{candyapplered}{rgb}{1.0, 0.03, 0.0}
\definecolor{ao(english)}{rgb}{0.0, 0.5, 0.0}
\definecolor{applegreen}{rgb}{0.55, 0.71, 0.0}

\definecolor{mygreen}{HTML}{1B9E77}
\definecolor{myblue}{HTML}{377EB8}
\definecolor{myblack}{HTML}{000000}

\usepackage{amsthm}

\usepackage{soul} 

\newcommand{\bcp}{\texttt{bcp}}
\newcommand{\tbf}{\boldsymbol{\tau}}

\newcommand{\Vbf}{\textbf{V}}

\newcommand{\score}{\Phi}
\newcommand{\pen}{\mathrm{pen}}

\newcommand{\rt}{r^{(t)}}
\newcommand{\norm}[1]{\left\|#1\right\|}

\newtheorem{assumption}{Assumption}

\newtheorem{definition}{Definition}

\DeclareMathOperator*{\argmax}{arg\,max}

\newcommand{\av}{\underline{a}}

\long\def\comment#1{}

\newcommand{\Rbb}{\ensuremath{\mathbb{R}}}

\newcommand{\Tt}{\mathbb{T}}

\newcommand{\prop}{\text{prop}}
\newcommand{\cur}{\text{cur}}
\newcommand{\new}{\text{new}}
\newcommand{\bd}{\text{bd}}
\newcommand{\shift}{\text{shift}}

\newcommand{\RSSr}{\widehat{\mathrm{RSS}}_r}
\newcommand{\one}{\mathds{1}}
\newcommand{\Mkr}{\mathbf{M}_{r^{\cur}}}
\newcommand{\dd}{\text{d}}

\usepackage{xr}

\newcounter{NumEx}[section]

\newcounter{deff}[section]
\renewcommand{\thedeff}{\arabic{section}.\arabic{deff}}

\newcounter{theormm}[section]
\renewcommand{\thetheormm}{\arabic{section}.\arabic{theormm}}

\makeatletter
\renewcommand{\paragraph}{\@startsection{paragraph}{4}{\z@}%
  {3.25ex \@plus1ex \@minus.2ex}%
  {-1em}%
  {\normalfont\normalsize\bfseries}}
\makeatother

\usepackage[font=small,labelfont=bf]{caption}

\sethlcolor{yellow}
\soulregister\ref7
\soulregister\eqref7
\soulregister\cite7

\numberwithin{equation}{section}

\begin{document}

\title{Change-in-velocity detection in multidimensional data}

\author{
Linh Do$^{1,*}$, Dat Do$^{2}$, Keisha J. Cook$^{3}$, Yitao Shen$^{4}$, and Scott A. McKinley$^{5}$\\[0.8em]
\small $^{1}$Department of Biostatistics and Bioinformatics, Duke University School of Medicine\\
\small $^{2}$Department of Statistics, University of Chicago\\
\small $^{3}$School of Mathematical and Statistical Sciences, Clemson University\\
\small $^{4}$Biochemistry and Molecular Biology and the Huck Institutes of the Life Sciences, The Pennsylvania State University\\
\small $^{5}$Department of Mathematics, Tulane University\\[0.5em]
\small $^{*}$Corresponding author: \texttt{thuylinh.do@duke.edu}
}
\date{}

\maketitle

\begin{abstract}
In this work, we introduce CPLASS
   (Continuous Piecewise-Linear Approximation via Stochastic Search), an algorithm for detecting changes in velocity within multidimensional data. The one-dimensional version of this problem is known as the change-in-slope problem (see \cite{cpop1, BaranowskiFryzlewicz2019}). Unlike traditional changepoint detection methods that focus on changes in mean, detecting changes in velocity requires a specialized approach due to continuity constraints and parameter dependencies, which frustrate popular algorithms like binary segmentation and dynamic programming. To overcome these difficulties, we introduce a tailored penalty function to balance improvements in likelihood due to model complexity, and a Markov Chain Monte Carlo (MCMC)-based approach with tailored proposal mechanisms for efficient parameter exploration. Our method is particularly suited for analyzing intracellular transport data, where the multidimensional trajectories of microscale cargo are driven by teams of molecular motors that undergo complex biophysical transitions. To ensure biophysical realism in the results, we introduce a speed penalty that discourages overfitting of short noisy segments while maintaining consistency in the large-sample limit. Additionally, we introduce a summary statistic called the Cumulative Speed Allocation, which is robust with respect to idiosyncrasies of changepoint detection while maintaining the ability to discriminate between biophysically distinct populations. 

\end{abstract}

\vspace{0.5em}
\noindent\textbf{Keywords:} Changepoint detection; intracellular transport; Markov chain Monte Carlo (MCMC); multidimensional time series; penalty functions; trajectory analysis.
\vspace{1em}

\section{Introduction}
The intracellular transport of organelles and cargo is a fundamental requirement of cellular function. Molecular motors carry lysosomes, vesicles, and other microscale cargo along microtubule filaments, producing trajectories marked by periods of directed motion, thermal fluctuation, and near-stationary pausing \citep{gross2004hither, kunwar2011mechanical, Hancock2014}. Understanding how transport properties change over time, in particular, detecting when a particle transitions between distinct velocity states, is central to characterizing the biophysical behavior of motor proteins and their collective dynamics. These transitions manifest as changes in the velocity of particle position over time, making this a multidimensional generalization of what is known in statistics as the change-in-slope problem.

Changepoint detection has been studied for over sixty years across many scientific domains, including signal processing \citep{signal}, financial analysis \citep{Bai1998LRT, Frick2014}, genomics \citep{Futschik2014DNAsegmentation, Niu2012DNAcopynumber}, and environmental science \citep{environment}. The canonical formulation is detecting changes in the mean of independent observations, a problem for which many effective methods exist. However, the change-in-velocity problem poses a distinct set of challenges. Because particle positions are continuous in time, the increment process cannot be treated as simply changing in mean: continuity constraints couple parameters across segments in ways that frustrate the most widely used algorithms. Binary segmentation, for instance, iteratively searches for a single changepoint at a time; in a velocity-change setting, an initial estimate placed between two true changepoints cannot be corrected in subsequent iterations, leading to spurious detections (Figure~\ref{fig:bin-segl}). Dynamic programming approaches such as PELT \citep{PELT} and Optimal Partitioning \citep{Optimal_Partition} are similarly unsuitable because the continuity requirement introduces parameter dependencies that violate the independence assumptions these methods rely on \citep{cpop}. Taking first differences to convert velocity changes into mean changes is another natural idea, but \cite{cpop1} showed this approach discards information and performs poorly in practice.

Existing methods for the change-in-slope problem include Trend Filtering \citep{trendfilter}, the Narrowest-Over-Threshold (NOT) algorithm \citep{BaranowskiFryzlewicz2019}, CPOP \citep{cpop}, and the Narrowest Significance Pursuit (NSP) \citep{Fryzlewicz2023}. These methods represent meaningful advances, but they are designed for one-dimensional data. Extending them to the multidimensional setting is nontrivial, and intracellular transport data are inherently two- or three-dimensional.

In the study of intracellular transport, traditional summary statistics such as mean squared displacement (MSD) analysis are widely used \citep{Monnier2012, Manzo2015}, but MSD-based approaches average over entire trajectories and can miss short-lived state transitions. More recent single-particle methods have focused on detecting changes in diffusivity \citep{Persson2013, Karslake2020}, velocity \citep{Neumann2017, Jensen2021}, or both simultaneously \citep{Yin2018}. In previous work, our group \citep{Jensen2021, Nat} used the Bayesian Changepoint (\bcp) algorithm to segment particle paths, modeling segments as discontinuous piecewise-linear functions plus stationary noise. While effective for many purposes, the discontinuous model can miss short, fast motile segments that are biophysically meaningful (Figure~\ref{fig:con-vs-discon}). Enforcing continuity of the underlying trajectory is not merely a mathematical distinction: it reflects the physical reality that cargo position evolves continuously in time, and failure to impose this constraint leads to systemic blind spots in the analysis.

\begin{figure}[!htb]
    \centering
    \includegraphics[width=1\linewidth]{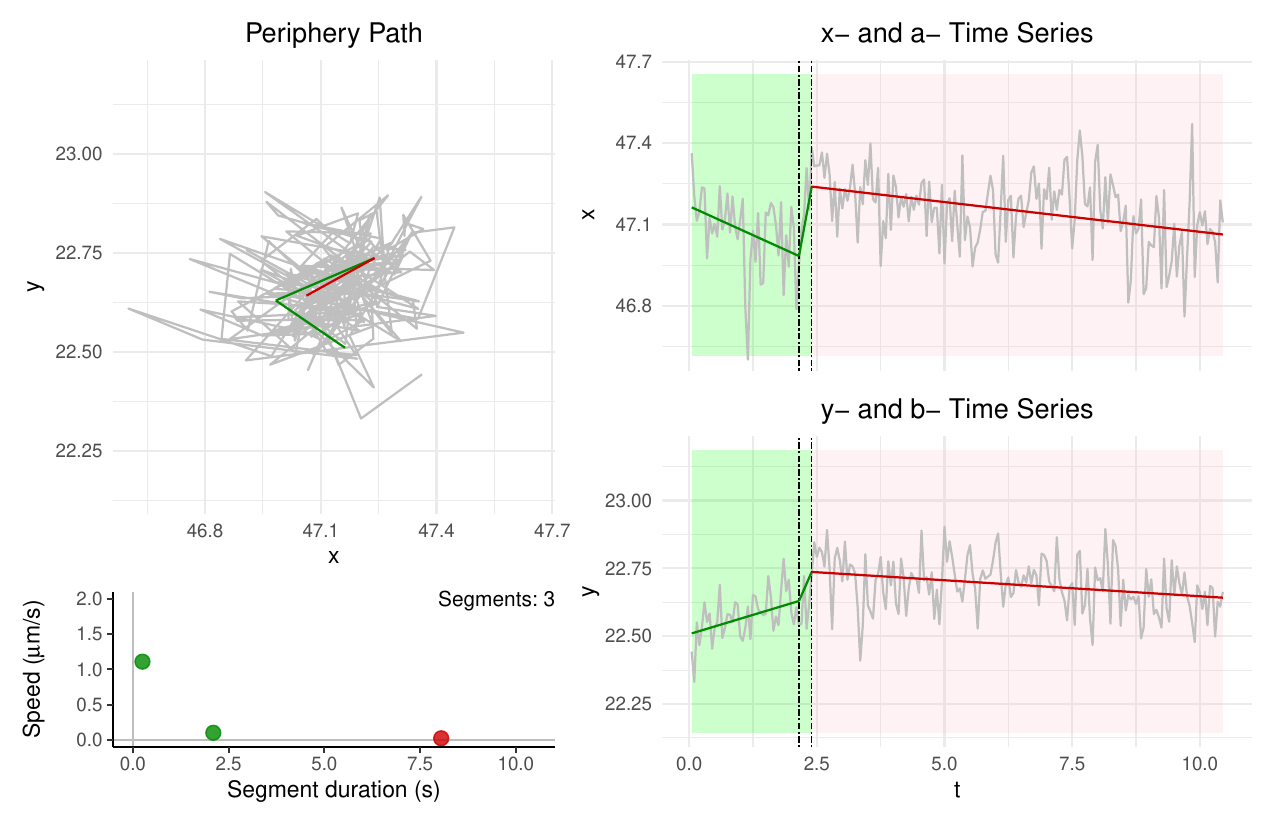}
    \caption[Example CPLASS segmentation of a peripheral lysosome trajectory]
{\textit{Representative lysosome trajectory in the periphery of a cell from \cite{Nat}.}
The \textcolor{gray}{gray} curves show the observed trajectory in the $x$--$y$ plane 
and the corresponding coordinate-wise time series. The \textcolor{mygreen}{green} and 
\textcolor{red}{red} curves denote the inferred anchor positions, colored by inferred 
state: \textcolor{mygreen}{green} indicates motile segments and \textcolor{red}{red} 
indicates stationary segments. \textcolor{mygreen}{Green} and \textcolor{red}{red} 
shaded regions indicate motile and stationary segments, respectively. In the 
segment-duration versus speed plot, point colors correspond to the inferred segment state.}
    \label{fig:example_PFpath34}
\end{figure}

\begin{figure}[!htb]
    \centering
    \includegraphics[width=1\linewidth]{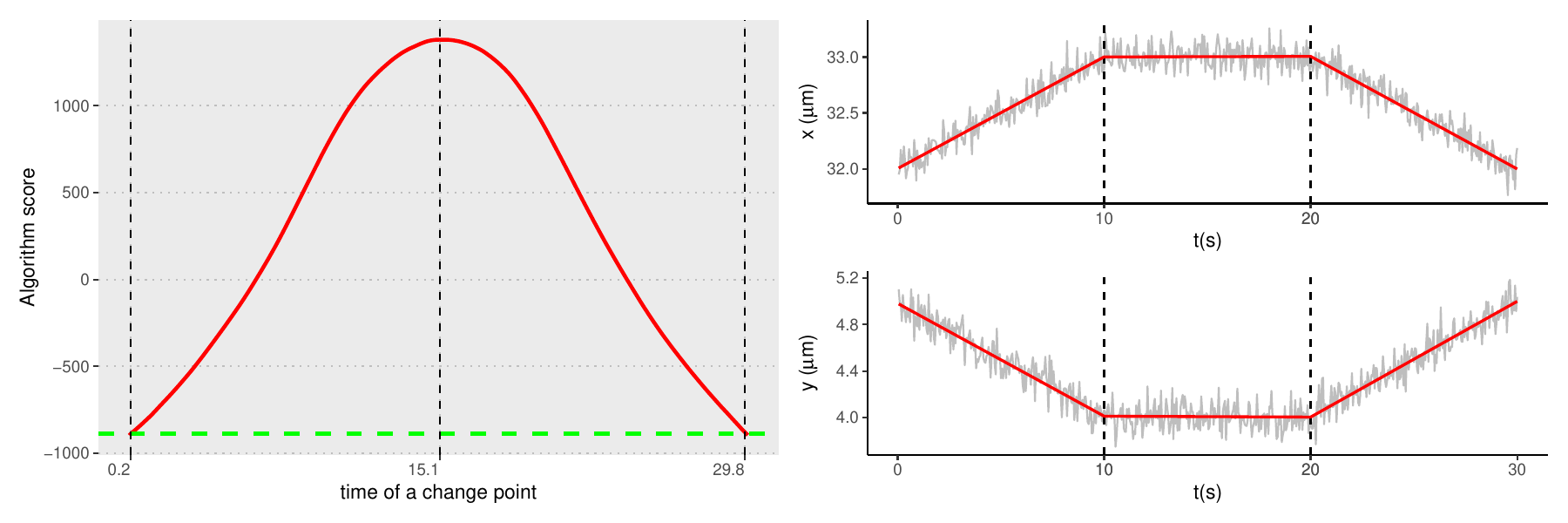}
    \caption[Binary segmentation failure for change-in-velocity detection]
{\textit{Failure of binary segmentation in a change-in-velocity setting}.
A 2D trajectory was simulated for 30\,s at 20\,Hz ($\sigma=0.1$), with true
changepoints at 10 and 20\,s and segment velocities
$(v_x,v_y)=(0.1,-0.1)$, $(0,0)$, and $(-0.1,0.1)\,\mu$m/s.
\textbf{Left}: CPLASS criterion values for the null (\textcolor{mygreen}{green dashed}) and
one-changepoint models (\textcolor{red}{red solid}), showing that binary segmentation
introduces a spurious changepoint (higher value is better).
\textbf{Right}: Observed $x$ and $y$ positions (\textcolor{gray}{gray}), true changepoints
(dashed), and fitted segmentation (\textcolor{red}{red}).}
    \label{fig:bin-segl}
\end{figure}

\begin{figure}[h]
    \centering
    \includegraphics[width=0.85\linewidth]{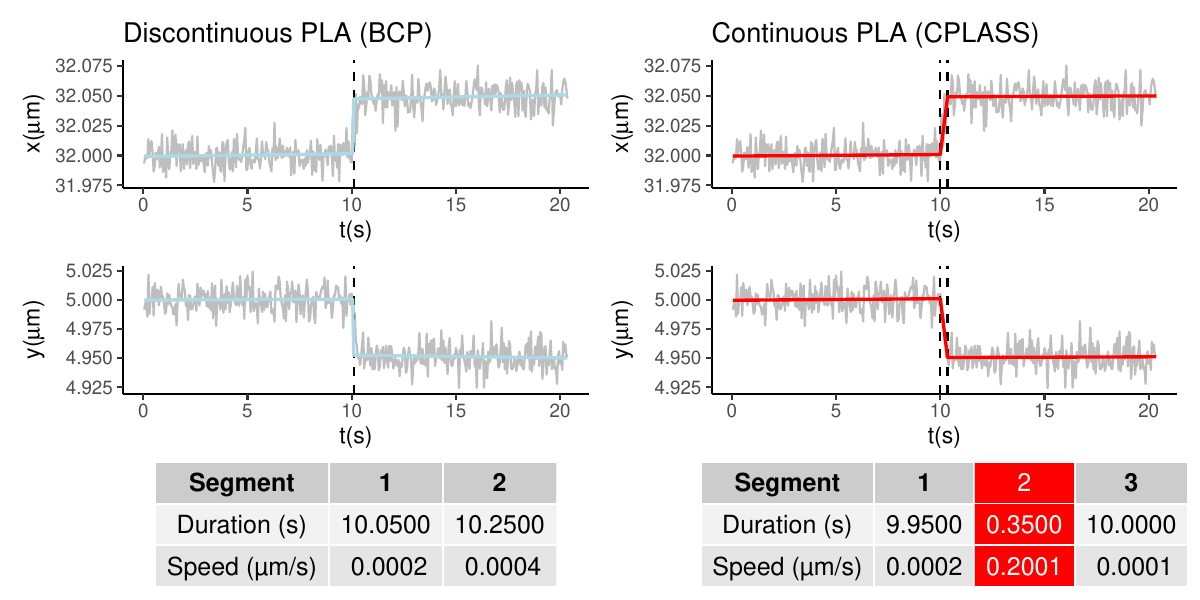}
   \caption[Discontinuous vs.\ continuous piecewise linear approximations]
{\textit{Discontinuous vs.\ continuous piecewise linear approximations}.
A simulated 2D lysosomal trajectory is shown via
$t$–$x$ and $t$–$y$ time series, with true changepoints at 9 and 10.35\,s.
Detected changepoints are shown as dashed lines, with the fitted
segmentation overlaid in \textcolor{red}{red}.
The red column in the CPLASS output table indicates a significant active
segment missed by BCP.
}
    \label{fig:con-vs-discon}
\end{figure}

To address these challenges, we introduce CPLASS (Continuous Piecewise-Linear Approximation via Stochastic Search), an algorithm for detecting changes in velocity in multidimensional particle trajectories. CPLASS models observed positions as independent Gaussian fluctuations around a continuous piecewise-linear anchor trajectory, and identifies the most likely segmentation of that anchor via a Markov Chain Monte Carlo (MCMC) search over the space of changepoint configurations. The criterion function combines a strengthened Schwarz Information Criterion (sSIC) penalty \citep{fryzlewicz2014segmentation, BaranowskiFryzlewicz2019} with a domain-informed speed penalty that discourages biophysically implausible segment speeds. We also introduce specialized proposal mechanisms, including a paired changepoint proposal, that are tailored to the structure of the change-in-velocity problem and substantially improve convergence when short, fast segments are present. Figure~\ref{fig:example_PFpath34} illustrates the CPLASS output for one experimental lysosome trajectory from the periphery of a cell. This example shows how CPLASS detects changes in velocity by estimating a continuous anchor path and partitioning the observed trajectory into distinct segments. For visualization, we label segments with inferred speed greater than 100~nm/s as motile and all remaining segments as stationary; this threshold is not used by the segmentation algorithm itself. The figure provides a concrete example of the segmentation structure produced by CPLASS before we systematically evaluate the method on simulated and experimental data.

We demonstrate the effectiveness of CPLASS on simulated trajectories and three experimental datasets: lysosomal transport in epithelial cells \citep{Nat}; quantum dot transport driven by kinesin-1, dynein-dynactin-BicD2 (DDB), and kinesin-1/DDB motor pairs \citep{Jensen2021}, and a preliminary application to EB1-GFP microtubule endpoint trajectories in neurons \citep{Thyagarajan2022MicrotubulePolarity, Scanlon2025NucleationFeedback}. To summarize segmentation output in a way that is robust to the idiosyncrasies of changepoint detection, we use the Cumulative Speed Allocation (CSA) statistic \citep{cook2024considering}, which quantifies the proportion of time a particle spends at each speed and enables principled comparison across experimental groups without requiring an arbitrary motility threshold.

Our contributions are threefold. First, we introduce CPLASS, a continuity-preserving method for detecting velocity changepoints in multidimensional intracellular trajectories. Second, we incorporate a biologically motivated speed penalty to discourage segments that require implausible object velocities while still preserving short active transport events. Third, we demonstrate the utility of the method through simulated trajectories and experimental datasets, and summarize inferred transport behavior using the Cumulative Speed Allocation statistic.


\paragraph{Data sets.} 
We used the following data sets to validate the proposed method. Throughout the paper, when we state that trajectories are observed or simulated at a given frequency (e.g., 20Hz or 25Hz), this means that the time step between consecutive observations is the inverse of the sampling frequency (e.g., 0.05s or 0.04s, respectively). The data include: (1) BS-C-1 monkey kidney epithelial cells and A549 human lung epithelial cells~\cite{Nat} (obtained from the Duke University Cell Culture Facility). For these data sets, intracellular transport is a requirement of cellular functions related to lysosomes; (2) data sets featuring quantum dots being transported by a single kinesin-1 motor, a single dynein-dynactin-BicD2 (DDB) motor, and a kin1-DDB pair~\citep{Jensen2021}, involving microtubule-based molecular motor transport; (3) MTEB1 microtubule endpoint tracking data, derived from live-cell imaging of neurites in studies of microtubule polarity~\cite{Scanlon2025NucleationFeedback, Thyagarajan2022MicrotubulePolarity}, consisting of two-dimensional particle trajectories from EB1-GFP comet tracking; (4) additional simulated data sets imitating trajectories in live-cell data~\citep{cook2024considering}.

\section{The CPLASS algorithm}\label{sec:the-cplass-algorithm}
In this section, we introduce the statistical model and algorithm used in this work.  In Section \ref{sec: statistical_model_CPLASS}, we introduce the statistical model.
In Section \ref{sec: PLA}, we construct a continuous piecewise-linear approximation of the data given the assumption that the number of changepoints is known. In Section \ref{sec: MCMC} and Section \ref{sec: MH as searching algorithm}, we propose a changepoint detection method where a criterion value with a penalty is provided, and a stochastic search method is used to find the maximum of the defined criterion function.
\subsection{Statistical model}\label{sec: statistical_model_CPLASS}

We assume that the $d$-dimensional observations
$\{Y_i\}_{i=1}^n \subset \Rbb^d$, collected at times
$\mathcal{T}:=\{t_1,\ldots,t_n\}\subseteq[0,T]$, are independent Gaussian
fluctuations around a sequence of unobserved anchor locations
$\{a_i\}_{i=1}^n\subset\Rbb^d$. We write
\begin{align}
    Y_i = a_i + \sigma\varepsilon_i,
\end{align}
where $\{\varepsilon_i\}_{i=1}^n$ is a sequence of independent and
identically distributed $d$-dimensional standard normal random variables,
and $\sigma>0$ is the noise magnitude. We use the conventions $t_0=0$ and
$a_0=\av\in\Rbb^d$.

Assume that the anchor trajectory has $k$ linear segments, with
$k+1\le n$, and let $\tau_j$ denote the time of the $j$th change in velocity,
for $j=1,\ldots,k-1$. For a natural number $k$, let
$[k]=\{1,\ldots,k\}$. We assume observations are made on a uniform grid of
size $\Delta=t_i-t_{i-1}=T/n$, for $i\in[n]$, and that each changepoint is
aligned with the observation grid, so that $\tau_j=t_{M_j}$ for some
integer $M_j\in\{1,\ldots,n-1\}$. We set
\(
  0=:\tau_0<\tau_1<\cdots<\tau_{k-1}<\tau_k:=T,
  M_0=0, M_k=n.
\)

Let $V_j\in\Rbb^d$ denote the velocity vector of the $j$th segment, with
speed
\(
  s_j=\|V_j\|_2, j\in[k],
\)
where $\|\cdot\|_2$ is the Euclidean norm. Within segment $j$, the anchor
locations satisfy
\begin{align}
    a_i = a_{M_{j-1}} + V_j(t_i-\tau_{j-1}),
    \label{eq:anchor_location}
\end{align}
for $i=M_{j-1}+1,\ldots,M_j$ and $j\in[k]$.

From the recursive formula \eqref{eq:anchor_location}, we construct a
multivariate continuous piecewise-linear signal function
$f_{\tbf,\Vbf,\av}:[0,T]\to\Rbb^d$, parametrized by changepoints
$\tbf=(\tau_1,\ldots,\tau_{k-1})$, velocity vectors
$\Vbf=(V_1,\ldots,V_k)$, and initial anchor position $\av\in\Rbb^d$. With
the convention $V_0=0$, the signal can be written as
\begin{equation}
    f_{\tbf,\Vbf,\av}(t)
    =
    \left(
    \av-\sum_{m=1}^{j}(V_m-V_{m-1})\tau_{m-1}
    \right)
    +V_jt,
    \qquad
    t\in[\tau_{j-1},\tau_j],\quad j\in[k].
    \label{eq:signal_function}
\end{equation}
Equivalently, using the hinge representation,
\begin{equation}
    f_{\tbf,\Vbf,\av}(t)
    =
    \av+V_1t+\sum_{j=1}^{k-1}(V_{j+1}-V_j)(t-\tau_j)_+,
    \qquad t\in[0,T],
    \label{eq:signal_function_hinge}
\end{equation}
where $(x)_+=x\mathbf{1}\{x>0\}$. 

When $V_{j+1}\neq V_j$ for all $j=1,\ldots,k-1$, the signal function
$f_{\tbf,\Vbf,\av}$ is said to have $k$ segments and $k-1$ changepoints.
Let $\mathcal{F}_k$ denote the collection of signal functions with $k$
segments.

We assume that the observations are generated from a true signal function:
\begin{align}\label{eq:true-model-1}
    Y_i \overset{\mathrm{ind.}}{\sim}
    \mathcal{N}\!\left(f^0(t_i),\sigma_0^2 I_d\right),
    \qquad i=1,\ldots,n,
\end{align}
where $f^0=f_{\tbf^0,\Vbf^0,\av^0}$ is the true signal function with
$k_0$ segments, true changepoints
$\tbf^0=(\tau_1^0,\ldots,\tau_{k_0-1}^0)$, true velocities
$\Vbf^0=(V_1^0,\ldots,V_{k_0}^0)$, and initial anchor position
$\av^0\in\Rbb^d$. Here $\sigma_0^2$ is the true variance and $I_d$ is the
$d$-dimensional identity matrix. Given the observations
$(Y_i)_{i=1}^n$, our goal is to infer the true number of segments $k_0$,
the signal parameters $(\tbf^0,\Vbf^0,\av^0)$, and the noise variance
$\sigma_0^2$.

Given a pre-specified upper bound $\overline{k}$ on the number of segments,
we estimate the parameters by maximizing the penalized likelihood:
\begin{align}\label{eq:fitted-pen-MLE_1}
    (\widehat{f}_{n}, \widehat{\sigma}_n^{2},\widehat{k}_n)
    =
    \argmax_{\substack{f\in \mathcal{F}_{k},\, \sigma^2 \in \Omega\\
                       k\leq \overline{k}}}
    \left\{
    \sum_{i=1}^n
    \log \mathcal{N}\!\left(y_i\mid f_{\tbf,\Vbf,\av}(t_i), \sigma^2 I_d\right)
    - \mathrm{pen}_k
    \right\},
\end{align}
where $\mathrm{pen}_k$ is a penalty term to be defined later. Equivalently,
for each $k\in[\overline{k}]$, we first compute
\begin{equation}\label{eq:each-MLE_1}
    (\widehat{f}_n^{(k)}, \widehat{\sigma}_{n,k}^{2})
    =
    \argmax_{f\in \mathcal{F}_k,\, \sigma^2 \in \Omega}
    \sum_{i=1}^{n}
    \log \mathcal{N}\!\left(y_i \mid f_{\tbf,\Vbf,\av}(t_i), \sigma^2 I_d\right),
\end{equation}
and then select
\begin{equation}\label{eq:pen_1}
     \widehat{k}_n
     =
     \argmax_{k\in [\overline{k}]}
     \left\{
     \sum_{i=1}^{n}
     \log \mathcal{N}\!\left(
     y_i \mid \widehat{f}_n^{(k)}(t_i),
     \widehat{\sigma}_{n,k}^2 I_d
     \right)
     - \mathrm{pen}_k
     \right\}.
\end{equation}
\subsection{Continuous piecewise-linear approximation given the changepoints}
\label{sec: PLA}

As noted above, maximizing \eqref{eq:fitted-pen-MLE_1} can be separated into
two steps. In this section, we address the first, which is to solve \eqref{eq:each-MLE_1}. That
is, given a set of fixed changepoints, we derive the maximum likelihood estimates of the continuous piecewise-linear model parameters.

Let $\mathbb{Y}=(y_{il})$ denote the $n\times d$ matrix of observed data.
Let $\mathbb{V}=(v_{jl})$ be the $k\times d$ matrix containing all segment
velocities, and let $\mathbb{W}=(w_{jl})$ be the $k\times d$ matrix of
velocity increments, defined by
\[
  \mathbb{W}[1,\cdot]=V_1,
  \qquad
  \mathbb{W}[j,\cdot]=V_j-V_{j-1},
  \quad j=2,\ldots,k.
\]
For fixed changepoints, we estimate the velocity increments
$w_{jl}$, the initial anchor position
$\av=(\av_1,\ldots,\av_d)$, and the noise magnitude $\sigma$.

Since the $d$ coordinates are conditionally independent given the signal,
we can estimate the regression coefficients separately in each dimension.
Let
\[
  Y^{(l)}=(y_{1l},y_{2l},\ldots,y_{nl})^\top\in\Rbb^n,
  \qquad l\in[d],
\]
be the $l$th column of $\mathbb{Y}$, and let
\[
  \underline{W}^{(l)}
  =
  (\av_l,w_{1l},\ldots,w_{kl})^\top
  \in\Rbb^{k+1}
\]
contain the $l$th intercept and velocity-increment coefficients. Then the
model in the $l$th dimension can be written as
\begin{align}
    Y^{(l)}=\Tt\underline{W}^{(l)}+\sigma \varepsilon^{(l)},
    \label{matrix form 1}
\end{align}
where
\begin{align}
\Tt
&=
\bigl[\,1,\ t,\ (t-\tau_1)_+,\ \ldots,\ (t-\tau_{k-1})_+\,\bigr]
\Big|_{t=t_i,\ i=1,\ldots,n},
\label{matrix_T}
\end{align}
with $(x)_+=x\,\mathbf{1}\{x>0\}$,
$\Tt\in\Rbb^{n\times(k+1)}$, and
$\varepsilon^{(l)}\sim\mathcal{N}(0,I_n)$.

The residual sum of squares is
\begin{align}
\mathrm{RSS}(Y,t;\underline{W})
:=
\sum_{l=1}^d
\left\|Y^{(l)}-\Tt\underline{W}^{(l)}\right\|_2^2.
\end{align}
The log-likelihood is
\begin{align}\label{loglikelihood}
    \mathcal{L}(f,\sigma)
    &=
    \sum_{i=1}^{n}
    \log \mathcal{N}\!\left(
    y_i \mid f_{\tbf,\Vbf,\av}(t_i), \sigma^2 I_d
    \right) \nonumber\\
    &=
    -\frac{nd}{2}\left[\log(2\pi)+\log(\sigma^2)\right]
    -\frac{1}{2\sigma^2}\mathrm{RSS}(Y,t;\underline{W}).
\end{align}
The resulting maximum likelihood estimators are
\begin{align}
    \widehat{\underline{W}}^{(l)}_{n,k}
    &=
    \left(\Tt^\top\Tt\right)^{-1}\Tt^\top Y^{(l)},
    \qquad l\in[d],
    \label{eq:W-hat}\\
    \widehat{\sigma}^2_{n,k}
    &=
    \frac{\mathrm{RSS}(Y,t;\widehat{\underline{W}}_{n,k})}{dn}.
    \label{eq: estimated sigma}
\end{align}
As long as the changepoints $\tau_j$ are distinct and the design matrix
$\Tt$ has full column rank, $\Tt^\top\Tt$ has a unique inverse. From now on,
we drop the subscript $(n,k)$ in the MLEs for ease of notation.

The velocity vector corresponding to the $j$th segment is recovered from the
velocity increments as
\[
  \widehat{V}_j
  =
  \left(
  \sum_{r=1}^{j}\widehat{w}_{r1},
  \ldots,
  \sum_{r=1}^{j}\widehat{w}_{rd}
  \right)^\top,
  \qquad j\in[k].
\]
The speed of the associated segment is
\[
  \widehat{s}_j=\|\widehat{V}_j\|_2.
\]
\subsection{Criterion function}\label{sec: MCMC}

The typical representation of the model contains changepoint times $\tbf$, initial intercepts (or initial anchor locations) $\av$, segment velocities $\Vbf$, and noise $\sigma$. In the following, we introduce a reduced representation of the model that uses changepoint vector $r=(r_1,...,r_{n-1})$ of ones and zeros, in the spirit of \citet{Lavielle}. When $r_i = 1$, this indicates that a changepoint has occurred during the $i$th time step.
Henceforth we use the notation $|r|$ to denote the number of changepoints (hence in the notation introduced in the last section, $|r| = k-1$). For any given a particular changepoint vector $r$, we can find an associated piecewise linear approximation whose $\mathrm{RSS}(Y,t;\widehat{\av}, \widehat{W}), \tau_{j}, \hat{s}_j$ ($j \in  [|r|+1]$) are determined as discussed in Section~\ref{sec: PLA}. To this end, we will use the subscript $r$ to indicate these relationships, i.e., $\RSSr = \mathrm{RSS}(Y,t;\widehat{\av}_r, \widehat{W}_r), \tau_{j}, \hat{s}_j, \widehat{\av}_r, \tau_{j,r}, \hat{s}_{j,r}, \widehat{V}_{j,r}$. Altogether, for a given $r$, according to the derivation in Section~\ref{sec: PLA}, the piecewise linear model yields a maximized value of the log-likelihood function when it has the form
\begin{align}\label{eq: aproxlikelihood}
\widehat{\mathcal{L}}_n= \log(L(Y,t;\widehat{\av}_r,\widehat{W}_r,\hat{\sigma}_r))
 &= \log\left(\dfrac{1}{(2\pi)^{nd/2}}\left(\dfrac{dn}{\RSSr}\right)^{nd/2} \exp\left(-\dfrac{nd}{2}\right)\right)\\
 &=-\dfrac{nd}{2}\log\left(\RSSr\right) + C,
\end{align} 
 where $C$ is a constant. We define the criterion function of the algorithm as follows.



\begin{definition}[Form of the criterion function]
\label{def: score}
    \begin{align}
         \textcolor{black}{\score(r) = 2 \widehat{\mathcal{L}}_n - \mathrm{pen}(r) = -nd\log\left(\RSSr\right) - \mathrm{pen}(r)}.  \label{eq: score_function}
     \end{align} 
 \end{definition}
 
Here, \textcolor{black}{$	\mathrm{pen}(r)$} refers to the penalty term designed to prevent overfitting. We used a strengthened Schwarz Information Criterion (sSIC) penalty expressed as \textcolor{black}{$(\log n)^{\gamma} \rho$} where $\rho$ is the number of parameters in the model and require $\gamma >1$ (see \cite{BaranowskiFryzlewicz2019, Piotr2014WBS}), and a speed-control penalty function to mitigate the occurrence of unrealistic speed values. 


\begin{definition}[Form of the penalty function] \label{def: Penalty}
\begin{align}
    \textcolor{black}{\pen(r) =} \textcolor{black}{(\log n)^{\gamma}} \textcolor{black}{\rho} + \textcolor{black}{\sum_{j=1}^{|r|+1}h(\hat{s}_{j,r}-s_{cap})}, \label{eq:pen}
\end{align}
where \textcolor{black}{$\gamma >1$}, \textcolor{black}{$\rho = |r|(d+1)+2d+1 = k(d+1)+d$} is the total number of parameters of the model,  \textcolor{black}{$|r|$} is the number of changepoints, $k = |r|+1$ is the number of segments, \textcolor{black}{$\hat{s}_{j,r} = \norm{\widehat{V}_{j}}_2$ ($j = 1,..., |r|+1$)} is the estimated segment speed, \textcolor{black}{$s_{cap}$}, decided by the practitioner, is the maximum speed that has no penalty, and \textcolor{black}{$h(s) = \max\{0,s\}$}.
    
\end{definition}
The speed control function is added based on prior scientific knowledge of
biologically realistic particle speeds. If no reliable upper bound is available,
one can set $h(s)=0$, in which case the penalty reduces to a linear $\ell_0$
penalty in the form of a strengthened Schwarz information criterion (sSIC)
\citep{Piotr2014WBS,BaranowskiFryzlewicz2019}. We remark that $\gamma = 1$
corresponds to the standard Schwarz/Bayesian Information Criterion (SIC/BIC)
penalty considered by \citet{YAO1988181} in the context of multiple changepoint
detection. In CPLASS, we adopt $\gamma > 1$ to preserve the formal consistency guarantees
developed in the author's doctoral dissertation \citep{Do2025CPLASS} and being
prepared in a companion methodological manuscript
\citep{DoDoMcKinley2026CPLASSTheory}. This strengthened
penalty has also been successfully used in wild binary segmentation
\citep{Piotr2014WBS} and the Narrowest-over-Threshold algorithm
\citep{BaranowskiFryzlewicz2019}. Based on empirical experiments, we recommend
$\gamma = 1.01$, which preserves the practical behavior of SIC while remaining
theoretically well supported across a broad class of signals.

\subsection{Stochastic search of the changepoint space}\label{sec: MH as searching algorithm}
\begin{figure}[!h]
    \centering
    \includegraphics[width=0.8\linewidth]{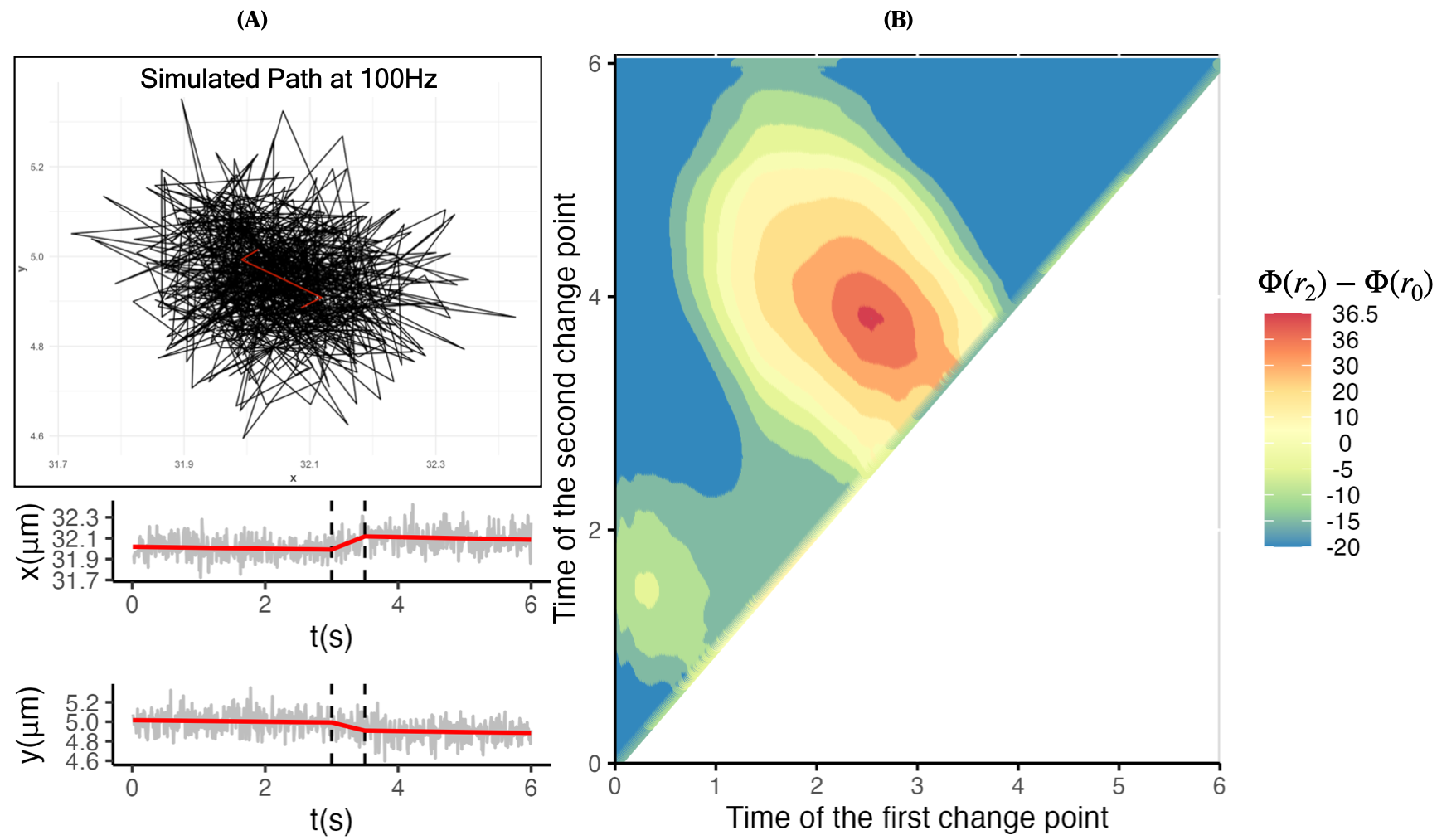}
    \caption[Why gradient ascent fails for change-in-speed detection]
{\textit{Example showing why gradient ascent is unsuitable for the change-in-speed problem}.
\textbf{(A)} A simulated 2D lysosomal trajectory (100\,Hz, 6\,s) with true
changepoints at 3 and 3.5\,s and segment speeds $(0,0.2,0)\,\mu$m/s.
Noisy $t$--$x$ and $t$--$y$ time series are shown in gray, with true
changepoints (dashed) and the true segmentation (red).
\textbf{(B)} Contours of CPLASS score differences between the two-changepoint
model $\score(r_2)$ and the no-changepoint model $\score(r_0)$.
Although the global maximum occurs at the true changepoints, gradient ascent
can become trapped in local maxima and fail to recover them.}

    \label{fig:gradient}
\end{figure}

When it comes to the search methods, the issue with using popular methods such as binary segmentation, PELT, or optimal partitioning has been discussed in the introduction. Another popular approach, called gradient ascent, has been used in optimization problems. However, this is not an ideal strategy because our criterion function commonly has multiple local maxima (see Figure~\ref{fig:gradient} for an example). We chose a stochastic search approach, using a Metropolis-Hastings algorithm as the search algorithm.

 \subsubsection{Metropolis-Hastings algorithm as the searching algorithm}

Notice that finding the maximum of the $\score(r)$ function is equivalent to finding the maximum of the function $
\exp(\score(r))$. Using Metropolis-Hastings, we can generate an ergodic Markov chain $\{\rt\}_{t\ge 0}$ that has $
C \exp(\score(r))$ as its stationary distribution. The maximum of the $\exp(\score(r))$ function can then be approximated by the maximum of the sequence $\{\exp(\score(\rt))\}$. It is important to keep in mind that if the proposal function is irreducible, then the Markov chain attained after running an MH algorithm is both irreducible and aperiodic. Moreover, since the chain $\{\rt\}_{t\ge 0}$ takes its values in a finite space, it is uniformly ergodic \citep{Lavielle2}. In Supplemental~3, we show that the detailed balance condition holds for our proposed rules.  
  
To be specific, in our algorithm, the proposal function for the changepoint process takes its values on the set $\{0,1\}^{n-1}$, where $n$ is the number of observations. There are four types of changepoint vector proposals: (1) an independent changepoint vector; (2) the creation or extinction of a changepoint; (3) the creation or extinction of a segment; or (4) a location shift of a single changepoint. Let $r^{\prop}$ and $r^{\cur}$ denote the proposed and current changepoint process, respectively. The following summary describes each type of proposal.
  \begin{itemize}
      \item \textbf{Type 1.} Notation $q_{\new}$. Including this type of proposal allows the changepoint space walk to escape local maxima, and for the number of changepoints to vary by proposing an independent change point vector, $r^{\prop}$ following the distribution of Bernoulli random variables with the probability of a changepoints is $1-\exp(-\lambda\Delta)$, where $\Delta$ is the time between observations: $r^{\prop}_i \overset{iid}{\sim} \mathrm{Bernoulli}(1-e^{-\lambda\Delta}), \text{ for } i\in[n-1].$
      We have that:
      \begin{align*}
          q_{\new}(r^{\prop}|r^{\cur})&=q_{\new}(r^{\prop}) \\
          &= (1-\exp(-\lambda \Delta))^{|r^{\prop}|} \times \exp(-\lambda\Delta)^{n-|r^{\prop}|-1},
      \end{align*}
      where $\lambda$ is chosen by the practitioner to qualitatively match the number of changepoints that is expected in the data set. (Without calibration to realistic numbers of changes, proposals of this type will tend to always be rejected.)
      \item \textbf{Type 2.} Notation $q_{\bd}$. While type 1 allows us to independently draw a new changepoint vector $r^{\prop}$, the second type proposal, $q_{\bd}$, provides another way for the number of changepoints to vary from iteration to iteration given the current changepoint vector $r^{\cur}$. Instead of using the proposal function  $q_{\bd}$ mentioned in \citet{Green}, we modify it such that the new version of $q_{\bd}$ will add one or delete one changepoint on the current list of changepoints with equal probabilities. In particular, let $\mathbf{M}_{r^{\cur}}=\{M_1,...,M_{|r^{\cur}|}\}$ be the set of all current changepoint indices and $r^{\prop}=r^{\cur}$. There is a $50\%$ chance that a component is randomly sampled from the current change point indices, $s\sim \mathrm{Uniform}\left(\mathbf{M}_{r^{\cur}}\right)$, then let $r^{\prop}_s=0$. Otherwise, a component $s$ is drawn from the complement of the current change point indices set, $s\sim \mathrm{Uniform}\left(\mathbf{M}^c_{r^{\cur}} \right)$, then $r_s^{\prop}=1$. For such a proposal, we have that

$$q_{\text{bd}}(r^{\text{prop}} | r^{\text{cur}}) =
\begin{cases}
\displaystyle \frac{1}{2|r^{\cur}|}\one_{\{r^{\cur} \neq \mathbf{0}\}}, & \text{if we remove a changepoint} \\
\displaystyle \frac{1}{2(n - 1 - |r^{\cur}|)}\one_{\{r^{\cur}\neq \mathbf{1}\}
      }, & \text{if we add a changepoint}
\end{cases}$$
      
      \item \textbf{Type 3.} Notation $q_{\bd_2}$. The third type of proposal, $q_{\bd_2}$, allows for adding or removing two nearby changes in the current list of changepoints. Like type 2, this proposal also provides a way to vary the number of changepoints in each iteration. In section \ref{numex: why_prop_type_3}, we provide motivation for including and removing consecutive changes. Let $r^{\prop} = r^{\cur}$. We set the chances of adding or deleting a segment to be equal.  In the case where we add a segment, a set $\{s,s'\} \subseteq \Mkr^c $ is randomly drawn from one of the $|r^{\cur
      }|+1$ segments $[M_{j-1}, M_{j}]$ ($\dd_{j} := M_{j}-M_{j-1}$) where $M_0 = 0, M_{|r^{\cur}|+1} = n,$ and $M_1, ..., M_{|r^{\cur}|} \in \Mkr$, then let $r^{\prop}_{s} = r^{\prop}_{s'} = 1$. In the case where we remove a segment, a set of two consecutive indices in the set of changepoints indices, i.e., $\{s,s+1\} \subseteq \Mkr$ ($|r^{\cur}| \ge 2$) is randomly chosen, we then let $r^{\prop}_s = r^{\prop}_{s+1} = 0 $. For this type of proposal, we have that
      
{\fontsize{9.5}{10}\selectfont
$$q_{\bd_2}(r^{\text{prop}} | r^{\text{cur}}) =
\begin{cases}
\dfrac{1}{|r^{\cur}|
      }\one_{\{|r^{\cur}| \ge 2\}
      }, & \text{if we delete a segment} \\
 \dfrac{1}{2}\sum_{j=1}^{|r^{\cur}|+1} \dfrac{(\dd_j-1)(\dd_j-2)}{(n-|r^{\cur}|-1)(n-|r^{\cur}|-2)},& \text{if we insert a new segment}
\end{cases}$$}

      where $\dd_j$ is the length of the $j$th segment and $\sum_{j}^{|r^{\cur}|+1}\dd_j = n$.
      
      \item \textbf{Type 4.} Notation $q_{\shift}$. This type of proposal allows exploration of the best combination of changepoints for a fixed number of changepoints. We obtain the proposed change point vector by randomly sampling two components of the current change point vector as follows:
     \begin{align*}
          s &\sim \mathrm{Uniform}\left(\mathbf{M}_{|r^{\cur}|}\right),\\
          s' &\sim \mathrm{Uniform}\left(\mathbf{M}^c_{|r^{\cur}|}\right).
      \end{align*}
      The proposal of this type, which is symmetric, is defined as
      \begin{align*}
         r_i^{\prop}=\left\{\begin{array}{@{}l@{}}
         r_i^{\cur},\hspace{1 cm} \text{ if } i \neq s,s'\\
    1-r_i^{\cur}, \hspace{0.4cm}\text{ otherwise. } 
  \end{array}\right.\, 
      \end{align*}
  \end{itemize}

We have that 
$$q_{\shift}(r^{\prop}|r^{\cur}) = \dfrac{1}{|r^{\cur}|}\times \dfrac{1}{n-1-|r^{\cur}|}.$$
  Finally, we combine all these proposal types to become one final proposal function:
  \begin{align}\label{eq: finalproposal}
         q_{r}(r^{\prop}|r^{\cur};u_r)=\left\{\begin{array}{@{}l@{}}
    q_{\new}(r^{\prop}), \hspace{0.9cm}\text{ if } 0\le u_r \le u_1\\
    q_{\bd}(r^{\prop}|r^{\cur}), \hspace{0.4cm}\text{ if } u_1<u_r\le u_2\\
    q_{\bd_2}(r^{\prop}|r^{\cur}), \hspace{0.3cm}\text{ if } u_2<u_r\le u_3\\
    q_{\shift}(r^{\prop}|r^{\cur}),\hspace{0.2cm}\text{ if } u_3<u_r\le 1,
  \end{array}\right.\, 
      \end{align}
    where $u_1, u_2 - u_1, u_3 - u_2, 1-u_3
    $ are probabilities that the proposal types 1, 2, 3, and 4 are chosen, respectively.
The sampling algorithm is then described in Algorithm \ref{alg:Gibbs}. We set $u_1= 1/4, u_2= 3/8, u_3 =1/2$ as default in the algorithm. We then introduce the CPLASS algorithm as in Algorithm \ref{alg:CPLASS}.

\begin{algorithm}[!htb]
\caption{MH algorithm: Unknown number of changepoints}\label{alg:Gibbs}
 {\fontsize{11}{10}\selectfont
 \hspace*{\algorithmicindent} \textbf{Input:} The observed data ($\mathbf{x},\mathbf{y},\mathbf{t}$), the rate of change point processes ($\lambda$), a time step ($\Delta$). \\
 \hspace*{\algorithmicindent} The number of iterations ($T_{max}$).\\ 
\hspace*{\algorithmicindent} \textbf{Output:} A list contains $T_{max}$ change point vectors $\{\rt\}_{t=0}^{T_{max}}$ with their corresponding $\RSSr(\mathbf{x},\mathbf{y},\mathbf{t}; \rt)$, $\tau_{k,\rt}, \hat{s}_{k,\rt}$ for $k\in[K_{\rt}]$.

\begin{algorithmic}[1]
    \State $t=0$. Draw randomly $r^{(0)}$ from a $n-1$ i.i.d $\mathrm{Bernoulli}(1-e^{-\lambda\Delta})$
    \State Compute $\RSSr(\mathbf{x},\mathbf{y},\mathbf{t}; r^{(0)})$, $\tau_{k,r^{(0)}}, \hat{s}_{k,r^{(0)}}$ (for $k\in[K_{r^{(0)}}]$) using the piecewise linear approximations. 
    
    \For {$t = 1, 2, \dots, T_{max}$}
        \State Draw $u_r \sim \mathrm{Uniform}(0,1)$; $r^{\prop}= q_r(.|r^{(t-1)};u_r)$ \eqref{eq: finalproposal}
        \State Compute the acceptance probability
        {\fontsize{10}{10}\selectfont \begin{align*}     
        \log\left(\alpha\left( r^{(t-1)},r^{\prop}\right)\right)=\begin{cases}
      \min\left\{0, \score(r^{\prop})-\score(r^{(t-1)})+\log\left(\dfrac{q_r(r^{(t-1)}|r^{\prop};u_r)}{q_r(r^{\prop}|r^{(t-1)};u_r)}\right)\right\}, &e^{\score(r^{(t-1)})} q_r(r^{\prop}|r^{(t-1)};u_r)>0\\
      0, &e^{\score(r^{(t-1)})} q_r(r^{\prop}|r^{(t-1)};u_r)=0.
    \end{cases}
        \end{align*}}
        \If{$\alpha\left( r^{(t-1)},r^{\prop}\right)\ge\mathrm{Uniform}(0,1)$}
         \State Set $\rt=r^{\prop}$
        \Else
        \State Set $\rt=r^{(t-1)}$
        \EndIf
        \State Compute $\RSSr(\mathbf{x},\mathbf{y},\mathbf{t}; \rt)$, $\tau_{k,\rt}, \hat{s}_{k,\rt}$ (for $k\in[K_{\rt}]$) by using the piecewise linear approximations.
    \EndFor
    \end{algorithmic}}
\end{algorithm}

\begin{algorithm}[ht!]
\caption{CPLASS algorithm}\label{alg:CPLASS}
 {\fontsize{11}{10}\selectfont \hspace*{\algorithmicindent} \textbf{Input:} The output from running Algorithm \ref{alg:Gibbs}. \\
\hspace*{\algorithmicindent} \textbf{Output:} A list contains the continuous piecewise linear approximation of the data in terms of $\mathbf{x}$ and $\mathbf{y}$, changes in time, segment durations, segment speeds.}
\begin{algorithmic}[1]
    \State Finding the maximum of the collected $\{\score(\rt)\}_{t= 1,...,T_{max}}$ and returning the corresponding $r^{(t*)}$.
    \State Using the continuous piecewise linear approximation with the finding $r^{(t*)}$ and returning the final output of the algorithm.
    \end{algorithmic}
\end{algorithm}


\subsubsection{Necessity of Type 3 proposals }
\label{numex: why_prop_type_3}

While it is natural to add, subtract, or shift individual points, the specific structure of the change-in-velocity model, which may involve biologically significant short, fast segments, also requires  proposals for adding or subtracting pairs of changepoints that would mark the endpoints of short segments. Certainly, MH will still succeed in the absence of such proposals, but finding short, fast segments can take much longer. This can happen because it is unlikely in long paths for the algorithm to randomly propose two positions that are close to each other, but there is a deeper issue that also impedes convergence. Similar to what is seen in Figure \ref{fig:bin-segl}, adding one endpoint of a short segment, but not the other, can lead to a decrease in the criterion function. So it becomes unlikely that a correct proposal of one short segment endpoint will be accepted without a companion endpoint. 

\begin{figure}[H]
    \centering
    \includegraphics[width=1\linewidth]{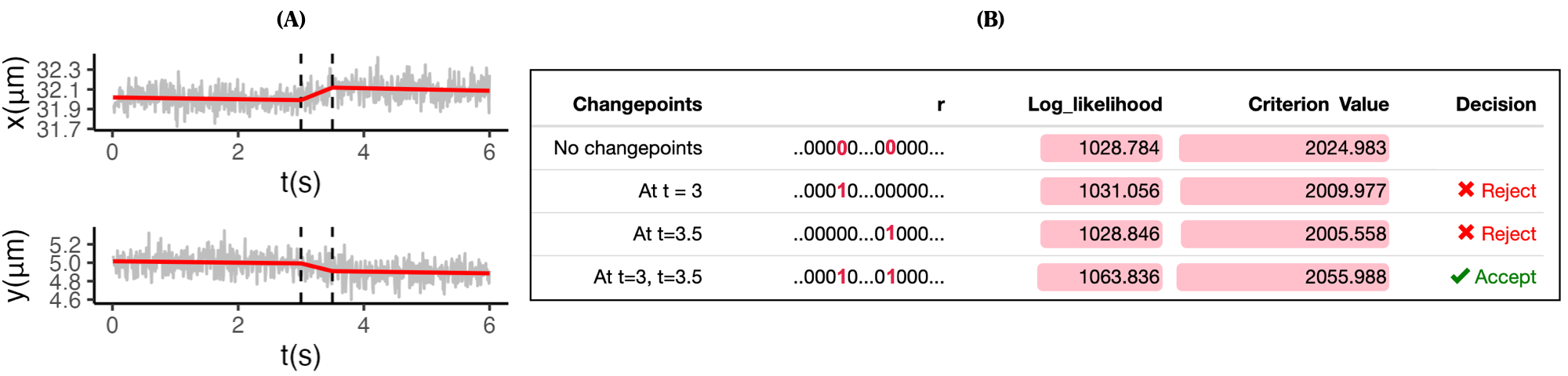}
   \caption[Necessity of the new proposal function]
{\textit{Necessity of the new proposal function} (Numerical Experiment~\ref{numex: why_prop_type_3}).
\textbf{(A)} A simulated 2D lysosomal trajectory (100\,Hz, 6\,s) with true changepoints at 3 and 3.5\,s,
shown in $t$--$x$ and $t$--$y$ coordinates. True changepoints are dashed; the fitted segmentation is overlaid in red.
\textbf{(B)} Candidate changepoint models with their log-likelihoods, criterion values, and corresponding
Metropolis--Hastings decisions (Algorithm~\ref{alg:Gibbs}).}
    \label{fig:why_add}
\end{figure}

We demonstrate the issue explicitly with the following numerical experiment. We simulated a 6-second path at 100Hz with changes at $t = 3$s and $t = 3.5$s. See Figure \ref{fig:why_add}(A). The segment speeds are $(0, 0.2, 0)$ $\mu$m/s, and we assume that the current changepoint vector is $r^{\cur} = \mathbf{0}$, i.~e., there are no changepoints. Figure \ref{fig:why_add}(B) provides the corresponding log-likelihood and criterion values for three proposals: two that involve adding one of the true endpoints of the active segment, and one that adds both simultaneously. Given the displayed criterion function evaluations, a progression from zero inferred changepoints to one and then two is prohibitively unlikely.

\section{Results}\label{sec:results}
This section evaluates the practical performance of CPLASS on biologically motivated simulations and on experimental intracellular transport datasets.

\subsection{Numerical experiments on the choice of the penalty term and comparison with BCP}\label{sec:needed_speed_pen}

{Now we validate the CPLASS algorithm and examine the penalty terms through simulation studies. 
Section~\ref{sec:vary_linear_pen} illustrates the performance of CPLASS when varying the value of $\gamma$ in the linear penalty term. Section~\ref{sec:nessity_of_the_speed_pen} demonstrates that CPLASS can detect short, fast segments better than BCP. However, in the presence of noise, the algorithm can sometimes detect a too-fast and too-short segment, leading to large speeds that are biophysically implausible. We will show that incorporating the speed penalty can help avoid this phenomenon (by slightly shifting the changepoint positions) without affecting the estimated number of changes or the overall signal function.
}
\subsubsection{Performance of CPLASS with different values of the linear penalty term}
\label{sec:vary_linear_pen}

\begin{figure}[!htb]
      \centering
      \subcaptionbox*{(A) \par} {\includegraphics[width = 0.49\textwidth]{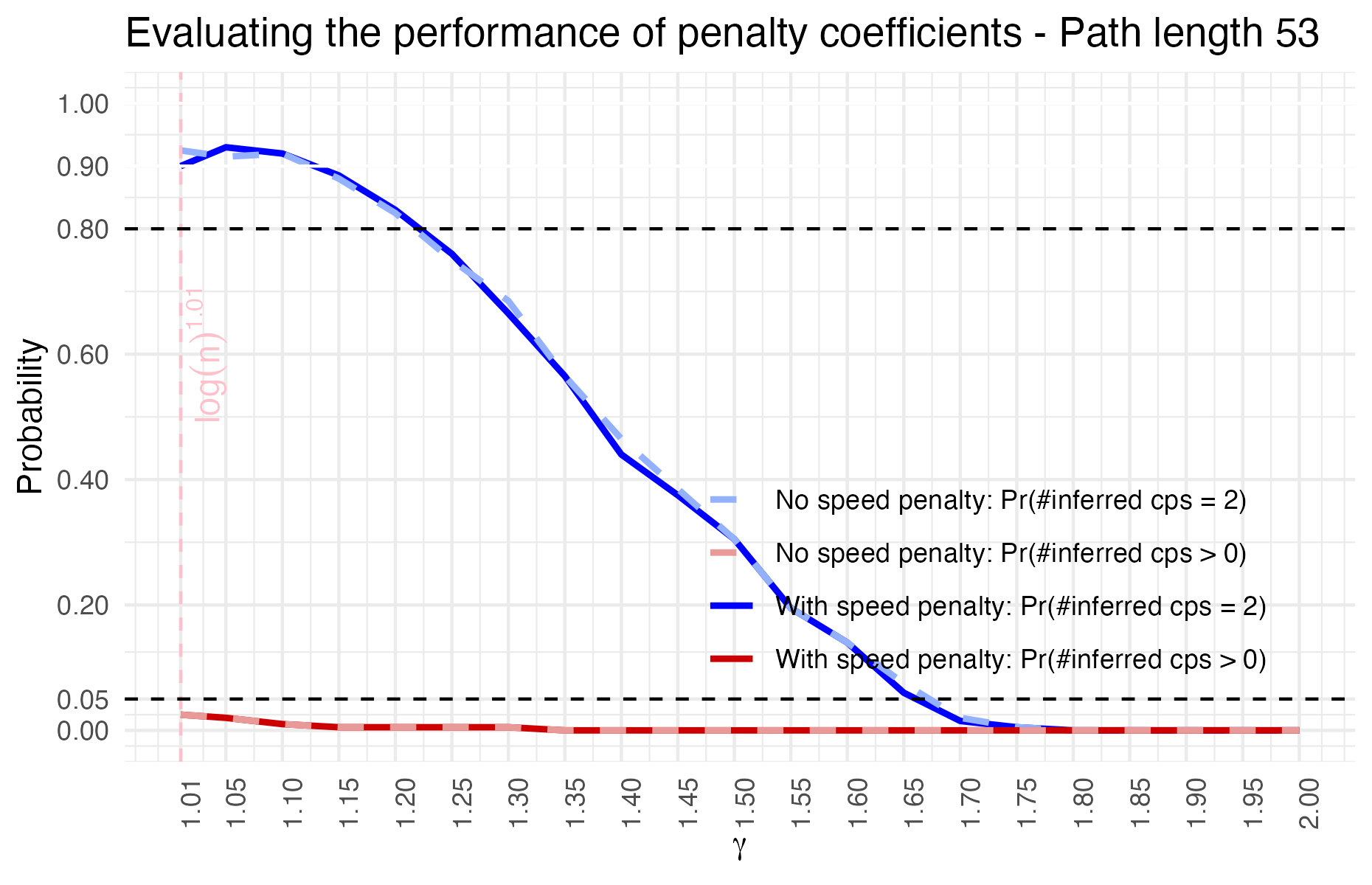}}
      \subcaptionbox*{(B)\par}{\includegraphics[width = 0.49\textwidth]{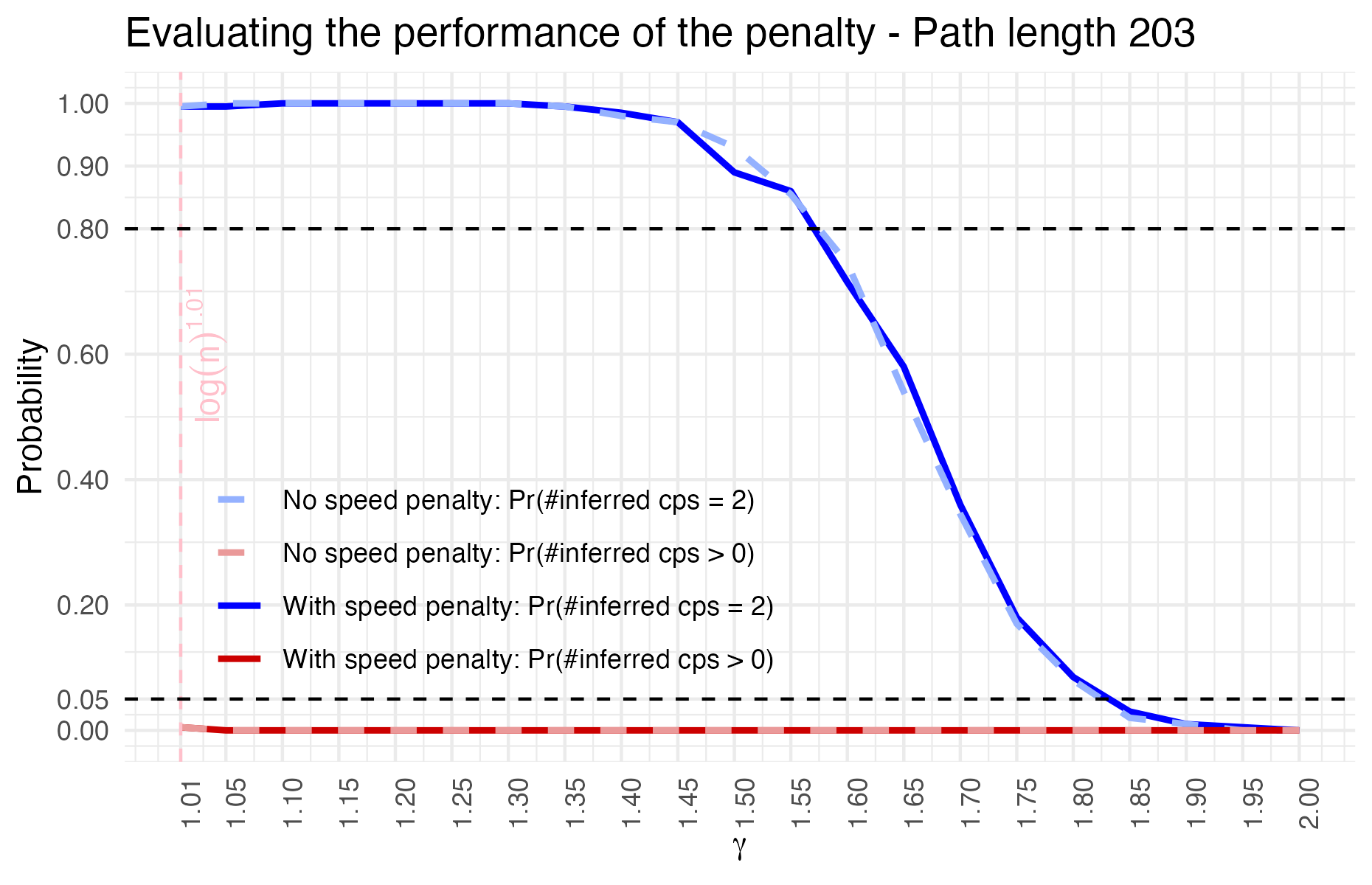}}
      \subcaptionbox*{(C)\par}{\includegraphics[width = 0.49\textwidth]{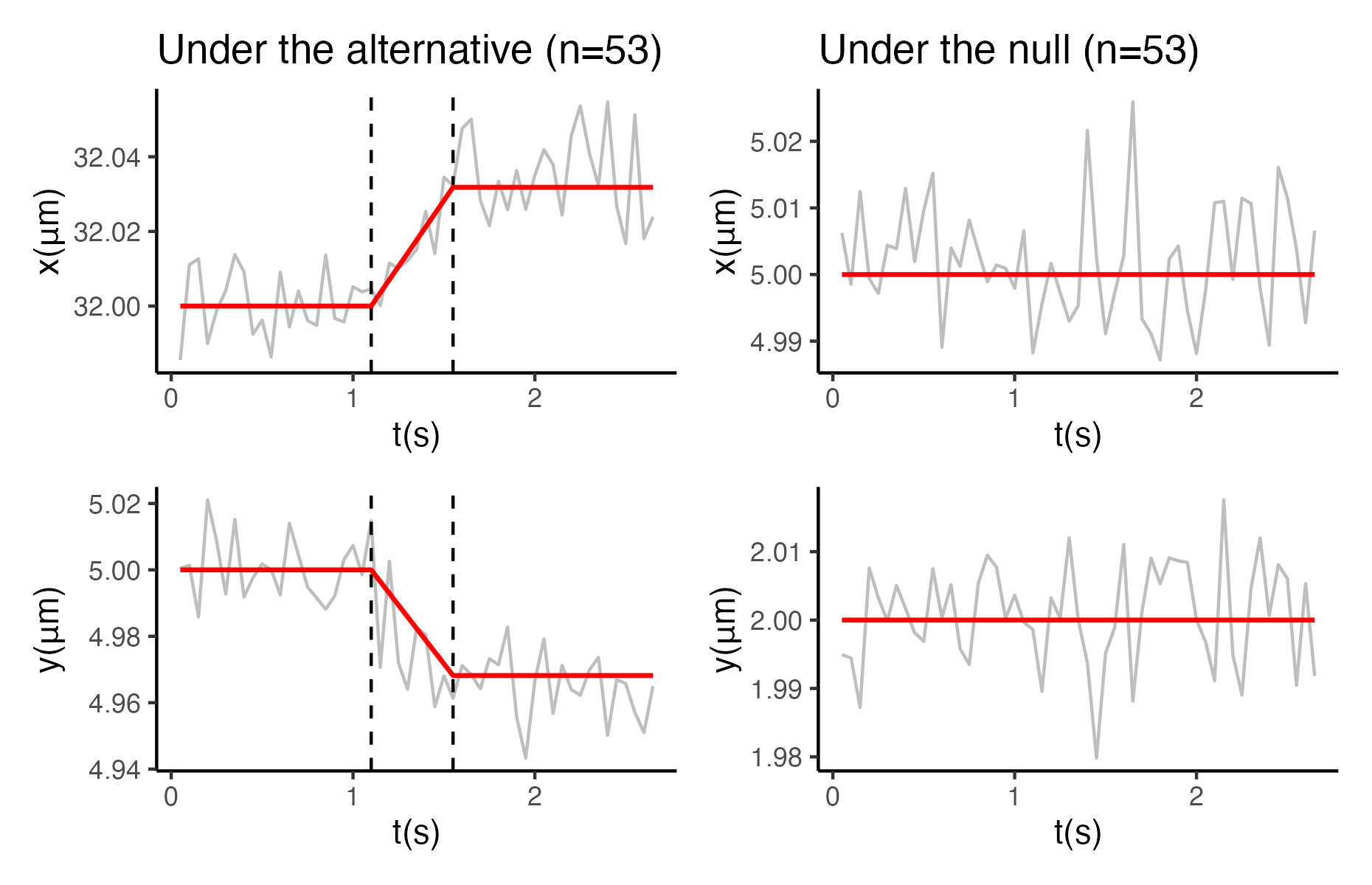}}
      \subcaptionbox*{(D)\par}{\includegraphics[width = 0.49\textwidth]{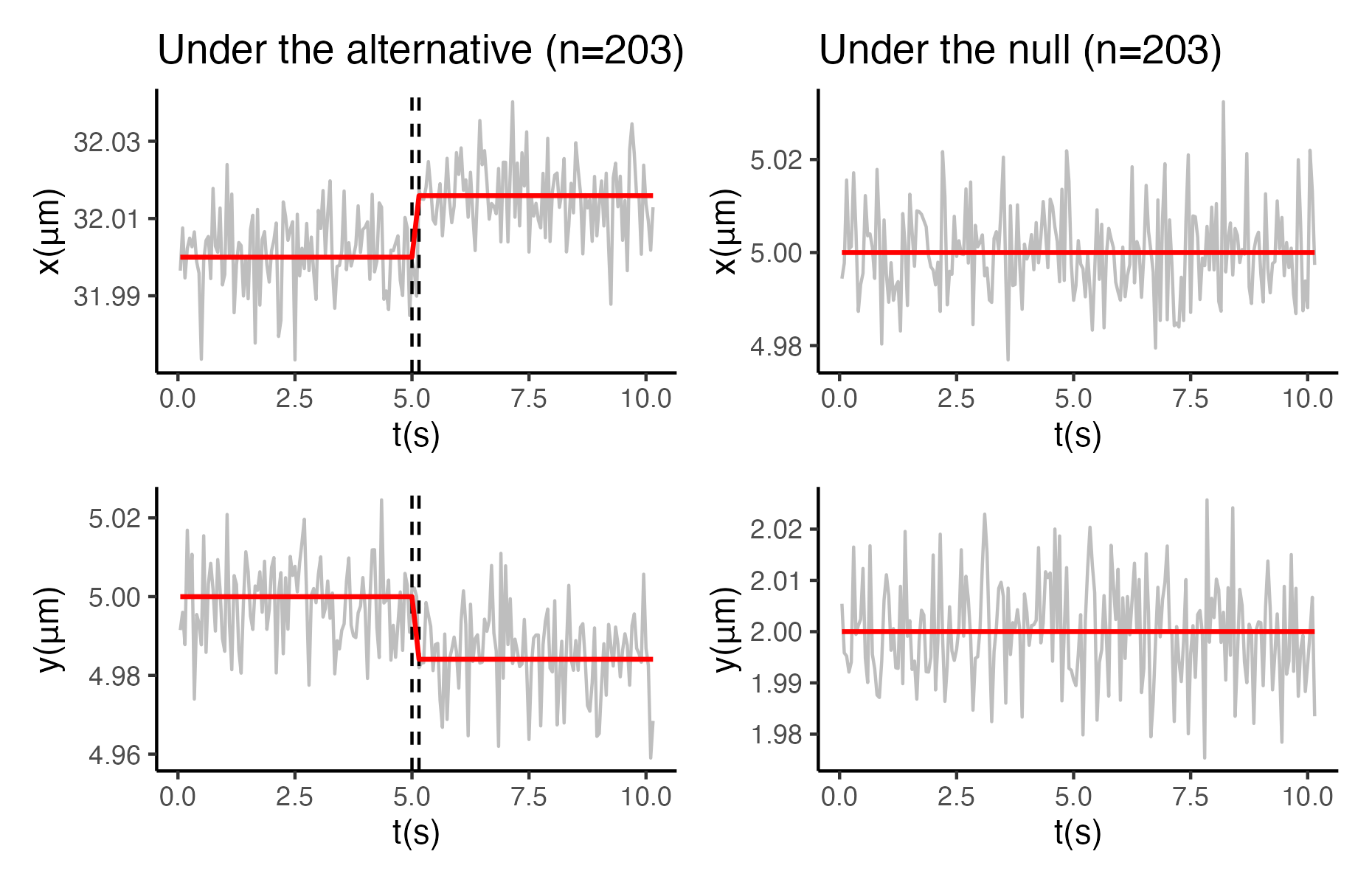}}
      
      \caption{\textit{Detecting short motile segments with CPLASS under varying $\gamma$
(Section~\ref{sec:vary_linear_pen}).}
\textbf{(A)} Detection probabilities for trajectories of length $n=53$
(20\,Hz, 2.65\,s). Under $H_0$, $s=0$; under $H_1$, two changepoints occur at
$t=1.1$ and $1.55$\,s with $(s_1,s_2,s_3)=(0,0.1,0)\,\mu$m/s ($\sigma=0.01$).
\textbf{(B)} Same analysis for $n=203$ (20\,Hz, 10.15\,s), with changepoints at
$t=5$ and $5.15$\,s and $(s_1,s_2,s_3)=(0,0.15,0)\,\mu$m/s.
\textcolor{blue}{Blue} curves show $\Pr\left(\widehat{|r|}=2\mid H_1\right)$ and \textcolor{red}{red} curves show
$\Pr\left(\widehat{|r|}>0\mid H_0\right)$ as functions of $\gamma$.
Solid (dashed) curves correspond to CPLASS with (without) the speed penalty.
\textbf{(C–D)} Representative trajectories under $H_1$ and $H_0$.}
\label{fig:vary_gamma}
\end{figure}

In this experiment, we evaluated CPLASS across different values of $\gamma$ and recommend
$\gamma = 1.01$ for reliable detection of \textit{short motile} segments when the sample size
$n \ge 50$ (Figure~\ref{fig:vary_gamma}). 
We considered the following models:
$H_0:\mathcal{M}_0 \text{ (no changepoints)},$ versus $
H_1:\mathcal{M}_1 \text{ (two changepoints)}.$ The simulation settings were chosen to challenge the algorithm by placing two changepoints
in close proximity (9 and 3 time steps for $n=53$ and $n=203$, respectively).
The speed of the middle segment was set to $0.1\,\mu$m/s for $n=53$ and
$0.15\,\mu$m/s for $n=203$. For \textbf{Panel (A)}, we generated two sets of 200 trajectories over $[0,2.65]$\,s at 20\,Hz ($n=53$).
Under $H_0$, paths were simulated with $\sigma=0.01$ and $s=0$.
Under $H_1$, two changepoints occurred at $t=1.1$\,s and $t=1.55$\,s with
$(s_1,s_2,s_3)=(0,0.1,0)\,\mu$m/s and $\sigma=0.01$.
CPLASS was run for $\gamma\in[1,2]$, with and without the speed penalty.
For each $\gamma$, we recorded the probability of correctly detecting two changepoints under $H_1$
and of incorrectly detecting changepoints under $H_0$. \textbf{Panel (B)} repeats this procedure for trajectories observed over $10.15$\,s at 20\,Hz ($n=203$).
Under $H_1$, changepoints occurred at $t=5$\,s and $t=5.15$\,s with
$(s_1,s_2,s_3)=(0,0.15,0)\,\mu$m/s.

As shown in Figure~\ref{fig:vary_gamma}, CPLASS correctly identified the true number of
changepoints in more than $90\%$ of trajectories across both settings, even when
changepoints were closely spaced and segment speeds were small.
Incorporating the speed penalty preserved high power under $H_1$ and keeping the
false detection rate under $H_0$ near zero.
We discuss the motivation for this penalty in the next section.

\subsubsection{Effect of the speed penalty}\label{sec:nessity_of_the_speed_pen}

\begin{figure}[t!]
      \centering
      \subcaptionbox*{(A) \hspace{7.5cm} (B)\par}{\includegraphics[width=1\linewidth]{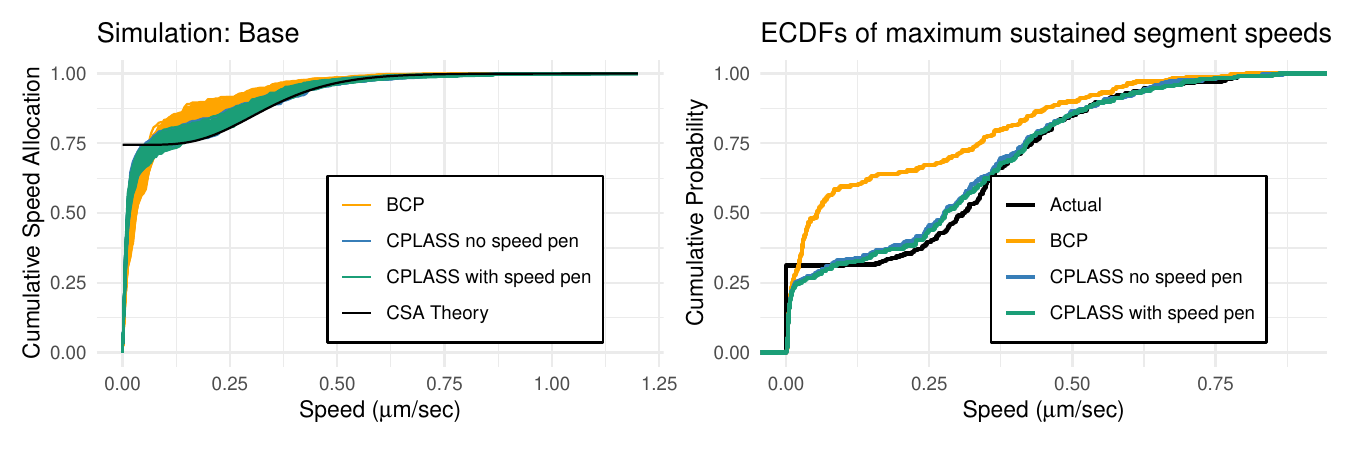}}
      \subcaptionbox*{(C) \par} {\includegraphics[width = 0.49\textwidth]{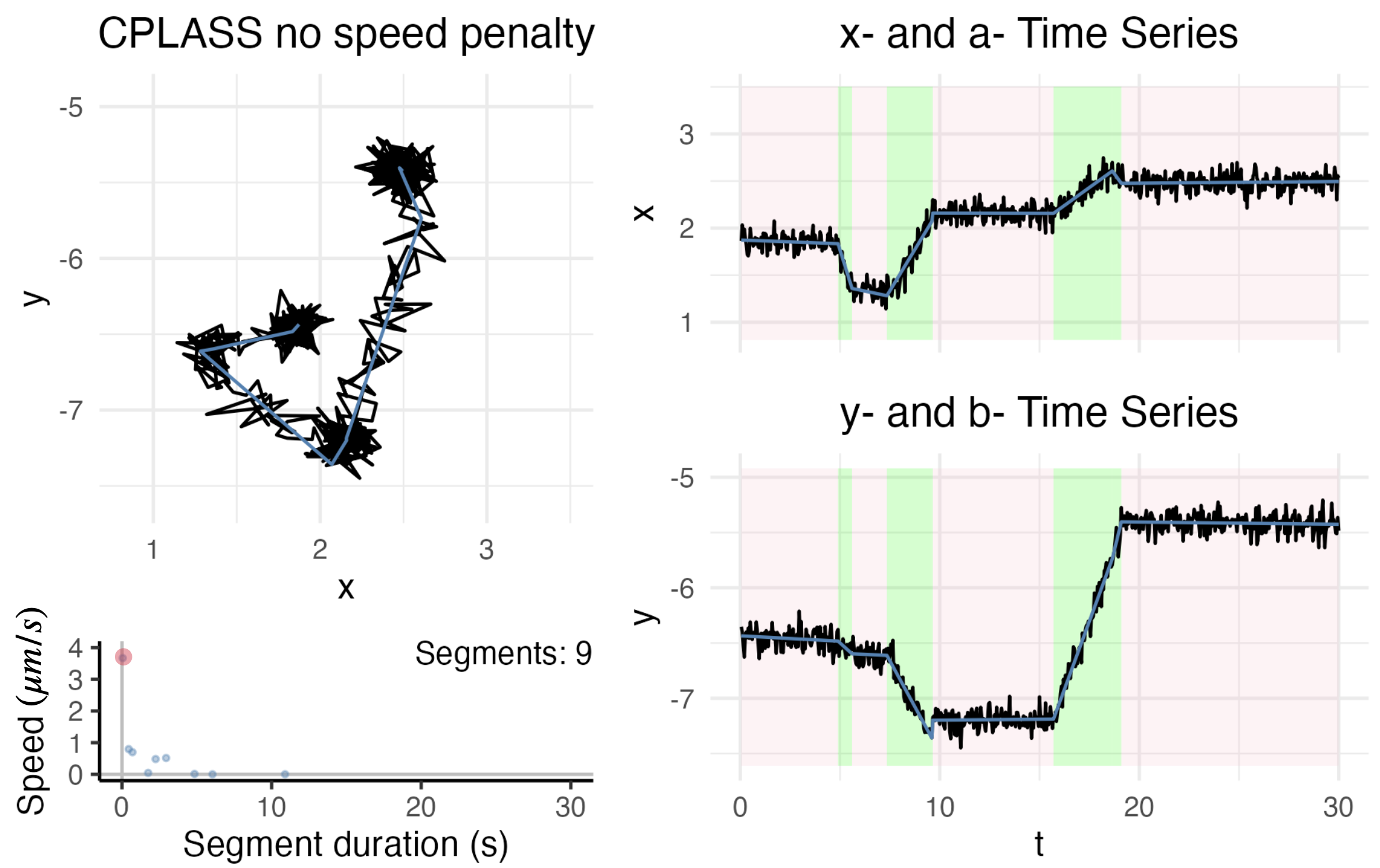}}
      \subcaptionbox*{(D)\par}{\includegraphics[width = 0.49\textwidth]{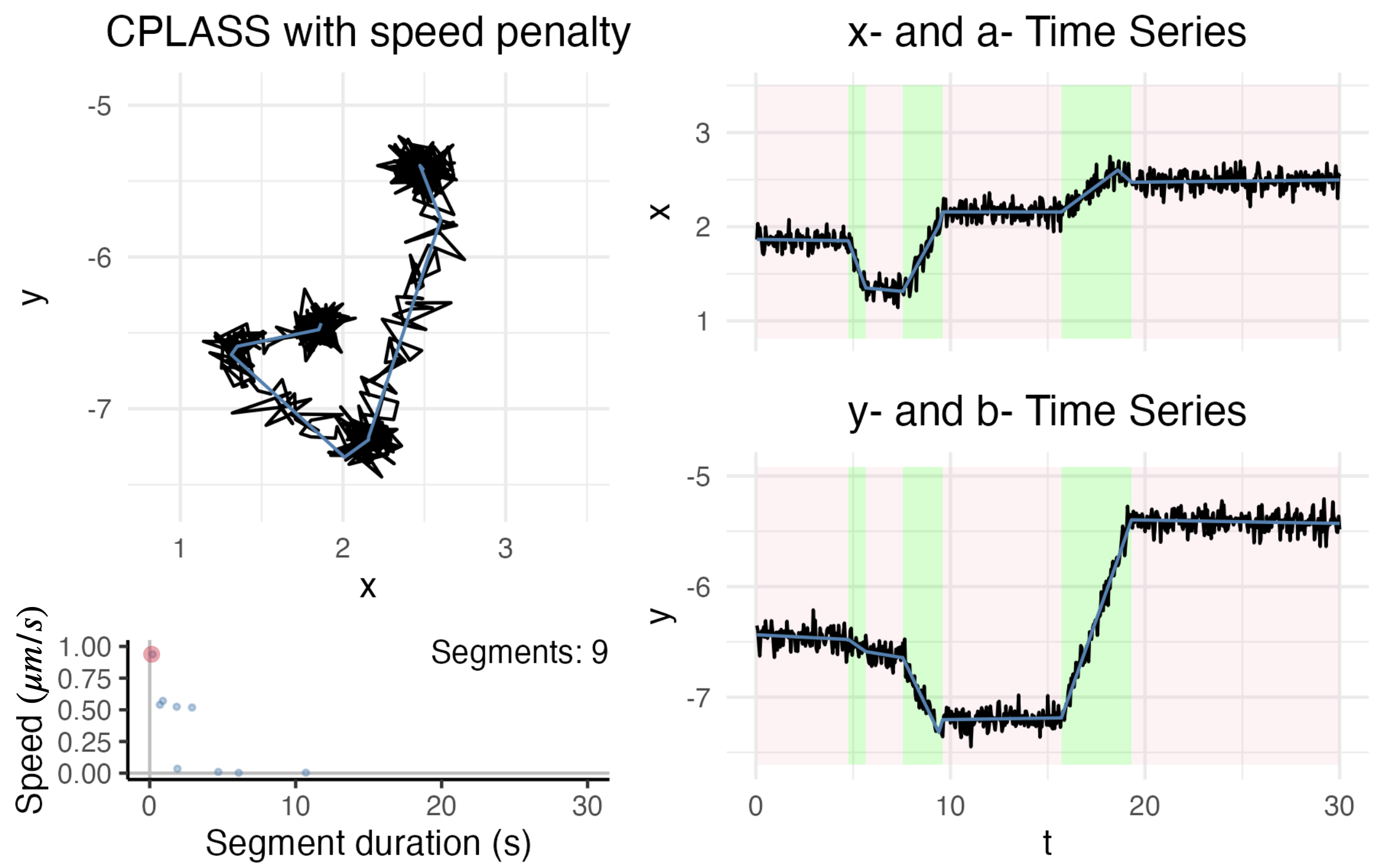}}\caption{\textit{Necessity of the speed penalty (Section~\ref{sec:nessity_of_the_speed_pen}).}
We simulated 250 trajectories from the parameter sets in Table~1 of \cite{cook2024considering}, sampled at 25\,Hz.
(\textbf{A}) Cumulative Speed Allocation (CSA): theoretical CSA (\textcolor{myblack}{\textbf{black}}),
CPLASS with speed penalty (\textcolor{mygreen}{\textbf{green}}),
CPLASS without speed penalty (\textcolor{myblue}{\textbf{blue}}),
and BCP \cite{Nat} (\textcolor{orange}{\textbf{orange}}).
(\textbf{B}) Empirical cumulative distributions of maximum sustained segment speeds
(minimum duration $0.6$s) for BCP (\textcolor{orange}{\textbf{orange}}),
CPLASS with (\textcolor{mygreen}{\textbf{green}}) and without
(\textcolor{myblue}{\textbf{blue}}) the speed penalty, compared with the true values
(\textcolor{myblack}{\textbf{black}}). (\textbf{C--D}) A representative trajectory after running CPLASS without/with speed penalty.
} \label{fig:CSA-CDF-base-simulation}
\end{figure}

We investigated CPLASS and BCP's performance on a collection of 250 simulated trajectories derived from the base parameter sets in Table 1 from \cite{cook2024considering} (also see Table~1 in Supplemental~2) at 25Hz, which is designed to mimic real trajectories. Recall that BCP models change-in-mean of location increments while CPLASS directly models the trajectories as piecewise-linear functions. We compared the inferred segmentation outputs by BCP, CPLASS (with and without speed penalties), and the truth by using two summary statistics, namely the Cumulative Speed Allocation statistic introduced in \cite{cook2024considering} and the Cumulative Distribution Function of the inferred maximum segment speeds (see Figure~\ref{fig:CSA-CDF-base-simulation}). Briefly, for every speed $s\ge 0$, the CSA is the inferred proportion of time spent at speeds less than or equal to $s$. We refer to \citet{cook2024considering} for a more detailed discussion on the CSA.

In Panel (A) of Figure~\ref{fig:CSA-CDF-base-simulation}, we display the result of applying the CPLASS (with and without the speed penalty) and BCP algorithms to 250 simulated trajectories. Each member of the CSA curve ensembles, orange for the BCP output, blue for CPLASS without the speed penalty, and green for CPLASS with the speed penalty, is the inferred CSA calculated from bootstrap resampling of the 250 paths. The evident gap between the CSA ensembles highlights the distinction between BCP and CPLASS, particularly in the proportion of time that the simulated particles are moving at speeds of $0.5\mu \text{m}/\text{s}$ or slower. Meanwhile, both versions of CPLASS (with the speed penalty activated or deactivated) closely follow the theoretical CSA curve (in black), which was used to simulate the data. This is consistent with the argument we discussed earlier in this paper regarding the issue of missing short fast segments when using the discontinuous piecewise linear model. 

The right panel of Figure~\ref{fig:CSA-CDF-base-simulation} illustrates the empirical cumulative distribution (ECDF) of the \textit{maximum sustained segment speeds} after running the BCP and the two versions of CPLASS. Here we present the ECDF of the maximum segment speeds where the associated durations are at least $0.6$s, since very short durations make it difficult to detect changepoints. We observed that CPLASS, with or without the speed penalty, more accurately matches the CDF of the true maximum speed than the results produced by BCP. {It is not immediately obvious from the ECDF or CSA that adding a speed penalty makes a difference, since both versions of the method approximate the true function equally well at the global level. However, differences become apparent in specific trajectories, such as those shown in Panels (C) and (D). In these cases, CPLASS with the speed penalty adjusts the estimated speed of short, fast segments from approximately $4\,\mu\mathrm{m}/\mathrm{s}$ to $1\,\mu\mathrm{m}/\mathrm{s}$ by slightly widening the estimated segment. Given that the speed limit $s_{\text{cap}} = 2\,\mu\mathrm{m}/\mathrm{s}$ is used here, we can see how the speed penalty softly regularizes the segment speed toward values below this threshold. 
}

\subsection{Change-in-mean versus change-in-velocity analysis}\label{sec:power_analysis}
Here we report the results of a direct power comparison between change-in-mean analysis of location increments (using BCP) versus continuous piecewise linear approximation (CPLASS with and without speed penalty) by simulating 20000 paths under the same conditions. All paths had two actual changes observed at 20Hz with $\sigma = 0.01$, but the positions of the changepoints varied due to differences in the duration and speed of the middle segment. The 20 values for the middle segment duration were selected over the interval from $0.05$ seconds to $1$ second, with an increment of $0.05$ seconds. The middle segment speed values (20 different values in $\mu$m/s) ranged from $0.01$ $\mu$m/s to $0.2$ $\mu$m/s, increasing by $0.01$ $\mu$m/s. We fixed the durations of the first and third segments (2 seconds) as well as those segment speeds (0 $\mu$m/s). Therefore, the sample sizes (number of location observations) varied from $n = 81$ to $100$. For each variation of the pair of speed and duration, there were $50$ corresponding simulated paths. The correctly detected percentage $\mathbf{P}_{\text{correct}}$ was then computed for each speed/duration pair as the proportion of paths in which the segmentation algorithm reported the correct number of changes. We ran CPLASS and BCP for these $20000$ simulation paths; the results showed that CPLASS (with both versions) was better than BCP in detecting short segments and slow segments, which BCP treated as having no movement (e.g., see Supplemental~1.2 Figure~3). Figure \ref{fig:power_analysis} illustrates the comparison. Regions in which the algorithms are considered "effective" are colored in green (see legend for color gradient). The region in which CPLASS (both with and without the speed penalty) is effective is notably larger than that of BCP. CPLASS also achieved better performance than BCP in detecting the true number of changepoints with $\mathbf{P}_{\text{correct}} \ge 90 \%$ for all cases where the middle segment had a duration varying from $0.45$s to $1$s and the speed varying from $0.08 \mu$m/s to $0.2 \mu$m/s.  Additionally, this experiment again confirms that adding the speed penalty function to the penalty function maintains the correctly detected percentage compared to using only the linear penalty term. 
    
\begin{figure}[t!]
    \centering
    \includegraphics[width=1\linewidth]{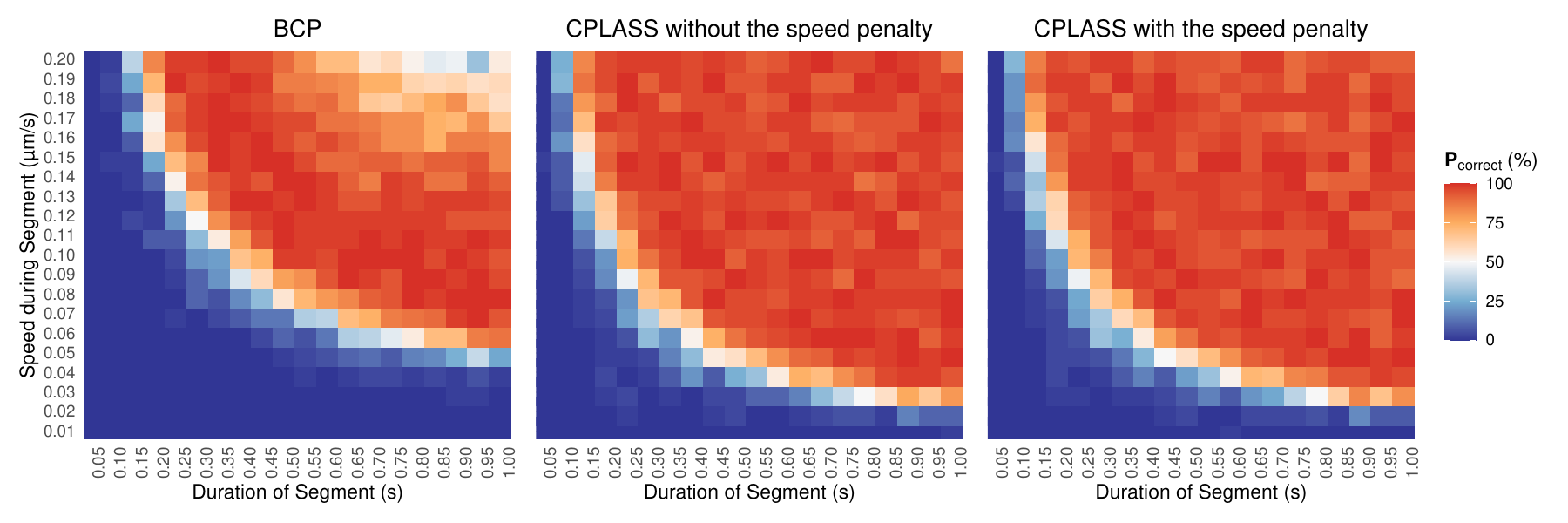}
    \caption{\textit{Power analysis comparing CPLASS and BCP
(Section~\ref{sec:power_analysis})}.
A total of 20{,}000 trajectories were simulated at 20\,Hz with two true
changepoints ($\sigma=0.01$).
The first and third segments have fixed durations of 2\,s, while the
middle segment varies across 20 durations and 20 speeds ($\mu$m/s).
For each speed--duration pair, 50 trajectories were generated.
The detection rate $\mathbf{P}_{\text{correct}}$ is the percentage of
paths for which the correct number of changepoints is identified.
Trajectories follow the same simulation design as
Figure~\ref{fig:vary_gamma}(C--D).}
\label{fig:power_analysis}
\end{figure}

\subsection{In vivo experimental data - lysosomal transport}\label{sec:vivo}
In this section, we revisit the data sets in \cite{Nat}. Two cell lines, monkey kidney epithelial cells, and human lung epithelial cells were cultured in different media but with identical conditions. Cells were supplemented with fetal bovine serum and incubated at $37^\circ$C and $5\%$ carbon dioxide. For imaging experiments, cells were transduced with CellLight Lysosomes-green emerald fluorescent protein to label lysosomes fluorescently. Transduction was carried out according to the manufacturer’s instructions. The study used live cell imaging and single-particle tracking to observe and characterize lysosome motion. A confocal microscope was used to collect images, and the TrackMate \citep{TINEVEZ201780} macro was used to track lysosomes. Lysosome trajectories were labeled as being in the perinuclear or peripheral regions of the cells, and were sorted by size. We refer to \cite{Nat} for more details about the data sets and data processing.

\begin{figure}
    \centering
    \includegraphics[width=1\linewidth]{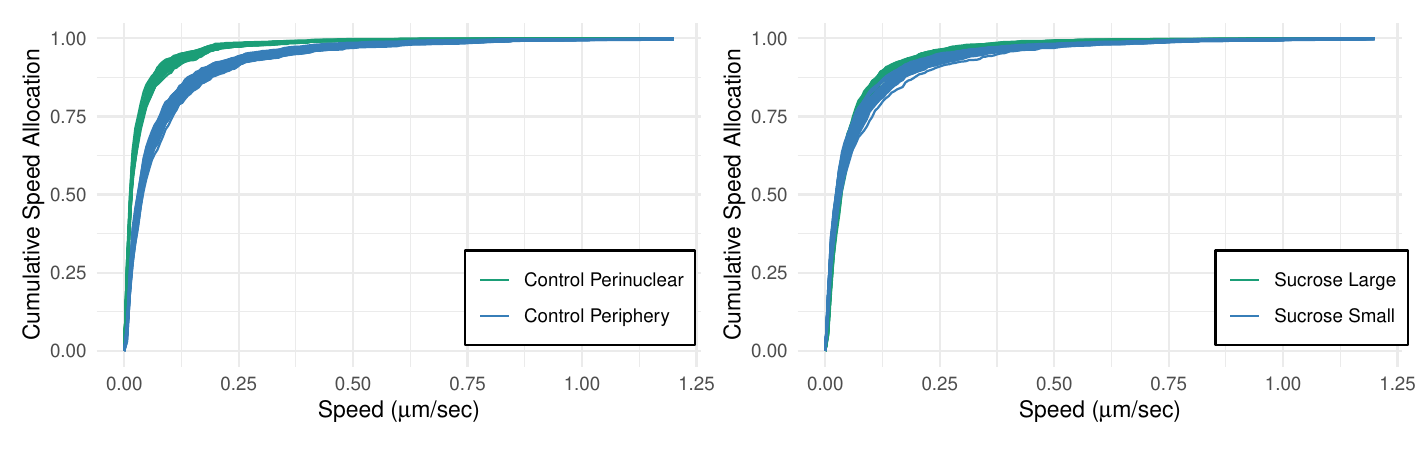}
    \caption{\textit{CSA plots for different group comparisons (Section~\ref{sec:vivo}).} (\textbf{Left}) The CSA bootstrap ensemble curves compare lysosomal transport in perinuclear and periphery regions. (\textbf{Right}) The CSA bootstrap ensemble curves compare the sucrose-treated groups restricted to the periphery region of the cell from~\cite{Nat}.}
    \label{fig:CSA_compare_PN_PF_PFL_PFS}
\end{figure}

We reassessed the queries regarding how lysosomal transport varies with lysosome size and location. Figure~\ref{fig:CSA_compare_PN_PF_PFL_PFS} indicates that intracellular location, rather than diameter, is a crucial factor in lysosomal motion. This aligns with the findings from \cite{Nat}. Analyzing the CSA plot (the left panel of Figure~\ref{fig:CSA_compare_PN_PF_PFL_PFS}), we observe that the lysosome in the perinuclear region spends more time moving slowly compared to that in the peripheral region.
The right panel of Figure~\ref{fig:CSA_compare_PN_PF_PFL_PFS} confirms that large lysosomes are slower in transport than small lysosomes; however, overall, there is not a significant difference between these two groups. In \cite{Nat}, to study the differences between these group comparisons, the authors first classify segment speeds into groups of motile ($s>0.1 \mu$m/s) and stationary ($s<0.1 \mu$m/s), then analyze the empirical cumulative distribution for the motile group. Meanwhile, using CPLASS and CSA, we can analyze the entire collection of segmented speeds and durations without establishing a threshold for the motile segment group.

\subsection{In vitro experimental data - quantum dot transport }\label{sec:vitro}
\begin{figure}[!htbp]
      \centering {\includegraphics[width = 0.95\textwidth]{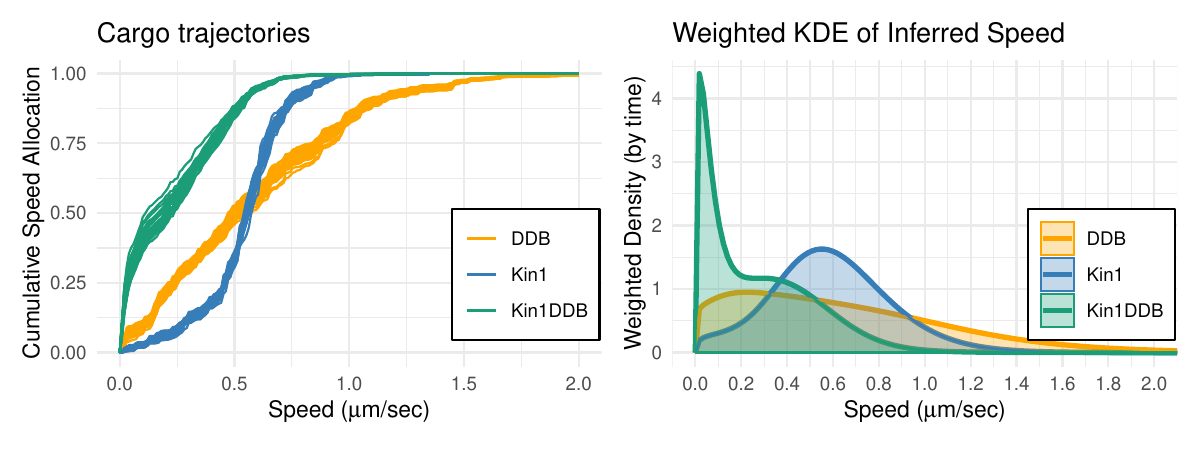}}
      \caption{\textit{CSA plot for different motor families (Section~\ref{sec:vitro}).} (\textbf{Left}) The CSA bootstrap ensemble curves compare cargo trajectories among three groups of motor transports: kinesin-1 (Kin1), Dynein-Dynactin-BicD2 (DDB), and Kinesin-1/DDB pairs. (\textbf{Right}) Weighted kernel density estimation of inferred segment speeds, duration-weighted, for the three groups.}\label{fig:will_hancock_CSA}
\end{figure}

This section revisited the data sets used in \cite{Jensen2021} in which quantum dots are transported by a single kinesin-1 (kin-1) motor, a single dynein-dynactin-BicD2 (DDB) motor, and by a Kinesin-1/DDB pair. In~\cite{Jensen2021}, the authors developed a protocol for finding changepoints in cargo trajectories that can be projected along the length of a straight microtubule and reporting velocity distributions. The differences in velocity distributions and run lengths revealed the behavioral difference for the biophysical state of three distinct motor-cargo arrangements. Since our proposed CPLASS can handle multidimensional data sets, we applied it directly to these quantum dot data sets without projecting the two-dimensional data into a one-dimensional format. We then calculated the CSA bootstrap ensemble curves (see the left panel of Figure~\ref{fig:will_hancock_CSA}) based on the collection of estimated segment speeds and durations in each motor-cargo system after running CPLASS. The CSA plot illustrates the differences among the motor experiments that correspond with what one might expect in a molecular motor ``folklore.'' In other words, Kinesin-1 motors step processively with consistent behavior, while DDB motors (orange curves) exhibit a broader range of speeds. When both motor types are present (green curves), the speed is generally lower, possibly reflecting a tug-of-war state. The right panel of Figure~\ref{fig:will_hancock_CSA} - weighted kernel density estimation (WKDE) of inferred segment speeds with weights given by segment durations across the three groups is a good tool for visualizing our argument. The tug-of-war phenomenon is more evident in this plot with more slow segments: one mode near $0$ (Kin1DDB), and the second mode is evident for cases when it appears one motor is dominant. The multimodality we observe in the WKDE here is also reported in the paper of \cite{Jensen2021}. All of these confirm the observations made about the data sets but offer a more refined and robust characterization. 

\subsection{Preliminary application to microtubule dynamics in neurons}
\label{sec:microtubule}

Microtubules constantly undergo polymerization and depolymerization within 
cells, providing both mechanical stability and transport infrastructure, 
particularly in long-lived cells such as neurons 
\citep{mahmood2020microtubule}. The growing plus ends of microtubules can be 
labeled with end-binding proteins such as EB1-GFP, producing fluorescent 
``comets'' whose trajectories provide a live-cell readout of microtubule 
polymerization dynamics \citep{Thyagarajan2022MicrotubulePolarity}. As a 
preliminary application, we applied CPLASS to EB1-GFP comet trajectories 
recorded in class~I dendritic arborization (da) sensory neurons in 
\textit{Drosophila} larvae.

The data analyzed here represent two videos, denoted top~2 and top~3, from a 
larger collection of 20 imaging sessions of the same cell type. The labels 
top~2 and top~3 refer to the orientation of the imaged dendritic field relative 
to the neuronal cell body, rather than to distinct experimental conditions. 
Full analysis of the complete dataset is ongoing. Comet tracking from raw 
fluorescence video requires substantial manual curation 
\citep{Thyagarajan2022MicrotubulePolarity}. TrackMate identified 47 candidate 
comet trajectories across the two videos; after excluding 14 short paths below 
the minimum observation count required for stable CPLASS fitting, 33 
trajectories remained for analysis (top~2: 16 trajectories, split across left 
and right sub-regions of the imaged dendrite; top~3: 17 trajectories). The 
imaging field of view was $34.97 \times 34.97\,\mu$m with a time increment of 
$1.79$\,s between frames.

Each trajectory was segmented by CPLASS into a sequence of piecewise-linear 
segments, producing estimates of local comet speed and segment duration. The 
number of inferred segments varied across tracks, reflecting heterogeneity in 
comet motion along different dendritic branches. In total, the filtered 
analysis produced 103 inferred segments over 1412\,s of tracked duration for 
top~2 and 86 inferred segments over 1377\,s for top~3. Inferred segment speeds 
were concentrated in the range of approximately $0.05$--$0.25\,\mu$m/s 
($3$--$15\,\mu$m/min), with most of the duration-weighted density centered near 
$5$--$7\,\mu$m/min. These values are broadly consistent with previously 
reported EB1 comet speed and run-length measurements in neuronal dendrites 
\citep{mahmood2020microtubule, Feng2019-en}.

\begin{figure}[!t]
\centering
{\includegraphics[width=0.95\textwidth]{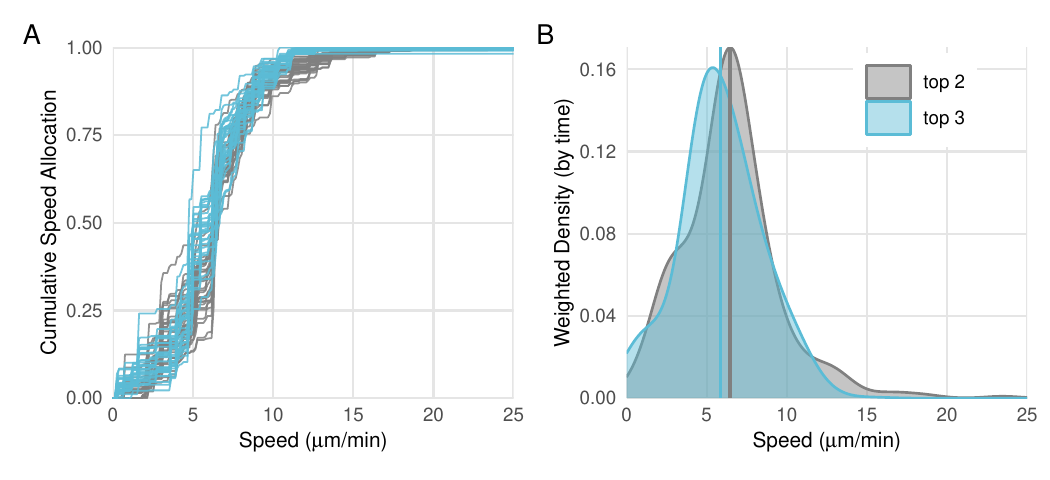}}
\caption{\textit{CSA and WKDE plots for EB1-GFP microtubule comet trajectories 
(Section~\ref{sec:microtubule}).} 
(\textbf{A}) The CSA bootstrap ensemble curves compare inferred speeds across 
two larvae of the same cell type (top~2: 16 trajectories, gray; top~3: 
17 trajectories, teal). 
(\textbf{B}) Duration-weighted kernel density estimate (WKDE) of inferred 
segment speeds for the same two animals. Vertical lines indicate the 
duration-weighted mean speed for each animal (top~2: $6.45\,\mu$m/min, gray; 
top~3: $5.87\,\mu$m/min, teal). Both panels report speeds in $\mu$m/min. 
The similar distributions across the two animals suggest that CPLASS provides 
a stable summary of EB1-GFP comet speed in this preliminary dataset.}
\label{fig:MT_CSA_WKDE}
\end{figure}

Figure~\ref{fig:MT_CSA_WKDE} summarizes the CPLASS output across both animals 
using the Cumulative Speed Allocation (CSA) and the duration-weighted kernel 
density estimate (WKDE) of inferred segment speeds, both reported in 
$\mu$m/min. The WKDE shows a unimodal, right-skewed distribution for both 
animals, with duration-weighted mean speeds of $6.45\,\mu$m/min for top~2 and 
$5.87\,\mu$m/min for top~3. The CSA curves show similar allocation patterns 
over the main speed range, indicating that the two preliminary datasets give 
comparable summaries of EB1-GFP comet motion. Occasional faster inferred 
segments are present, but they contribute relatively little to the 
duration-weighted distribution.

As in our lysosomal and quantum dot analyses 
(Sections~\ref{sec:vivo}--\ref{sec:vitro}), the CSA summarizes the full 
distribution of inferred segment speeds without requiring an arbitrary speed 
threshold to classify segments. In this setting, the segment-level output from 
CPLASS provides a compact way to summarize local variation in EB1-GFP comet 
motion along dendritic trajectories. In future work, once the full dataset has 
been processed, these summaries could be used to compare polymerization behavior 
across animals, dendritic regions, comet orientations, or genotypes.

We note one main limitation of the current analysis. CPLASS is applied here in
its standard piecewise-linear form, which does not yet explicitly account for
the curved and branched geometry of neuronal dendrites. A generalization to
non-linear piecewise paths suited to spatially complex neuronal geometries is
in active development.

Despite this limitation, these results demonstrate that CPLASS produces 
biologically plausible segmentations of EB1-GFP comet trajectories without 
requiring manual annotation of individual segments. The CSA statistic provides 
a stable preliminary summary for comparing comet-speed distributions across 
individuals and, in future work, across broader biological conditions.

\begin{figure}[!htbp]
\centering
{\includegraphics[width=0.95\textwidth]{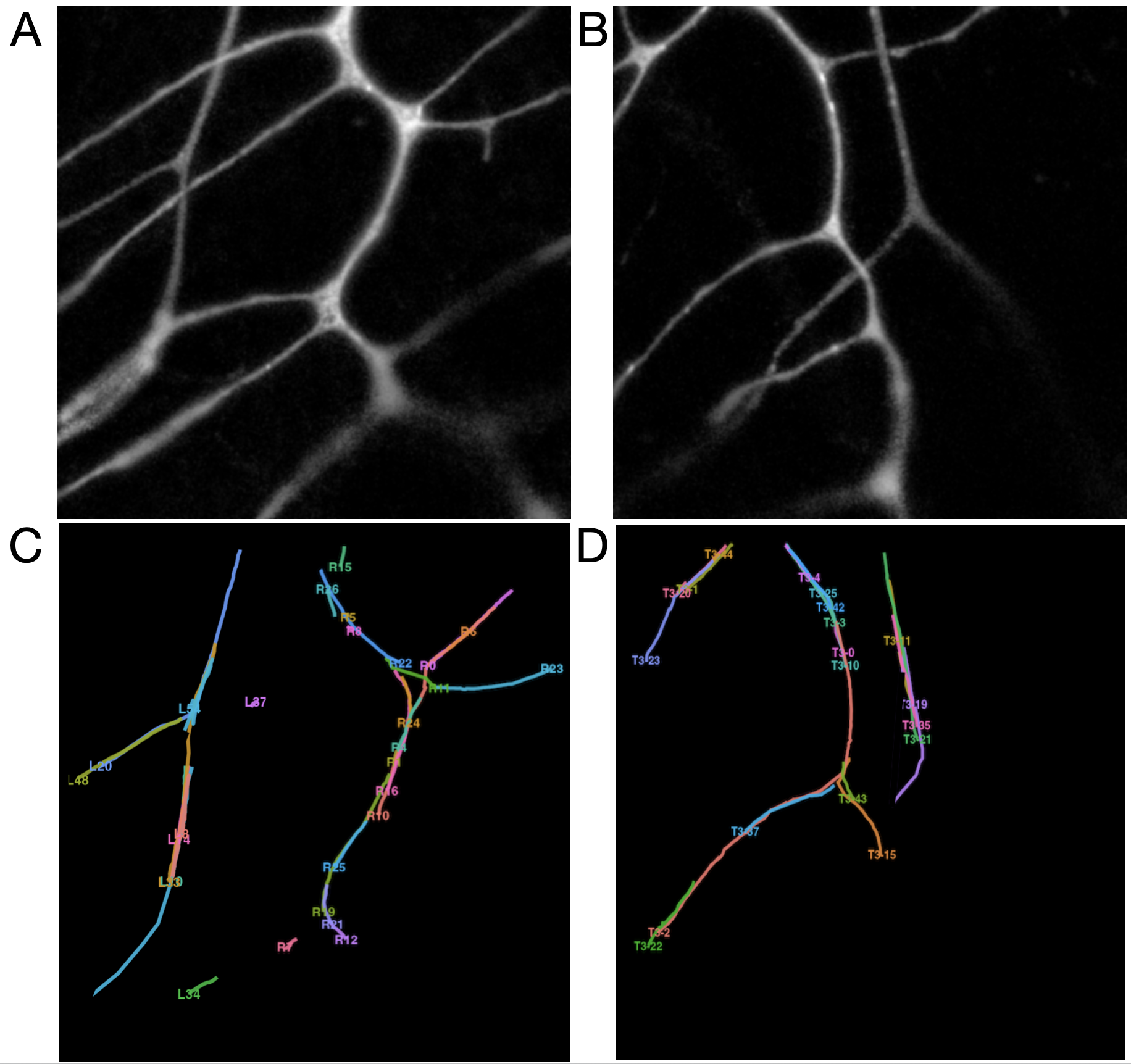}}
\caption{\textit{EB1-GFP microtubule comet imaging and TrackMate trajectory output (Section~\ref{sec:microtubule}).} (\textbf{A--B}) Representative first frames from two imaging videos of the same cell type (class~I da sensory neuron) from two different larvae of the same genotype (top~2 and top~3, respectively). Each video is a zoomed-in field of view of the dendritic arbor; the neuron's cell body (soma) lies outside the imaged region, above the field shown, and dendrite direction is defined relative to it (``top'' denotes the side nearer the soma, used to establish comet travel direction). EB1-GFP labels the growing plus ends of microtubules within these dendrites. The bright branching structures illustrate the curved and spatially complex geometry along which comets were tracked. (\textbf{C}) Comet trajectories extracted from the top~2 video using TrackMate, with each color representing a distinct tracked comet. (\textbf{D}) Comet trajectories for the top~3 video. These trajectories serve as input to CPLASS, which segments each path into piecewise-linear velocity segments as summarized in Figure~\ref{fig:MT_CSA_WKDE}. The branched and curved trajectory shapes in panels C--D reflect the underlying dendritic geometry visible in panels A--B.}
\label{fig:MT_data}
\end{figure}

\FloatBarrier

\section{Discussion}

 In this work, we introduced the CPLASS algorithm for detecting changes in velocity within multidimensional data, addressing key challenges in both probability structure and search methodology. While detecting changes in velocity seems to be a similar statistical problem to detecting changes in mean, it is fundamentally more challenging. Popular generic approaches for detecting multiple changepoints do not work for detecting changes in velocity. For example, popular ``bottom-up'' strategies, like binary segmentation, are initiated by assuming a single changepoint and finding the most likely location. However, as we discuss in the main text, the changepoint time inferred under the assumption that there is one change may not appear among (or near) the changepoints detected assuming there are two. This is a marked difference from the change-in-mean detection problem. Moreover, existing dynamic programming algorithms like PELT and optimal partitioning cannot handle change-in-slope due to continuity assumptions that create parameter dependencies, breaching essential independence structures. To address these issues, \citet{BaranowskiFryzlewicz2019} proposed the Narrowest-Over-Threshold (NOT) algorithm, while \citet{cpop} introduced a variant of dynamic programming, and \citet{trendfilter} offered trend-filtering methods. These work well for one-dimensional slope changes, but our challenge arises from analyzing multidimensional intracellular transport data. These challenges motivated our development of an MCMC-based approach, which includes specialized proposal mechanisms tailored to efficiently navigate the parameter space.

While consistency of the CPLASS estimator is established 
formally in our companion theoretical work 
\citep{DoDoMcKinley2026CPLASSTheory}, real-world applications, such as 
molecular motor data, pose additional challenges because 
practitioners often have prior knowledge of a practical upper 
bound on attainable speeds. To incorporate this information, 
we introduced a speed penalty that adjusts the placement of 
changepoints to discourage unrealistically short segments that 
would imply implausibly high speeds. This acts as a 
domain-informed prior, improving realism in the inference of 
speeds while preserving consistency in the large sample limit. 
Furthermore, by comparing Cumulative Speed Allocation (CSA) 
\citep{cook2024considering} with the Cumulative Distribution 
Function (CDF), we showed that quantifying the proportion of 
time spent at different speeds yields a more stable performance 
metric than segment velocity counts, making the approach less 
sensitive to algorithmic variation.

Crucially, our method is inherently multidimensional, allowing it to capture complex structures in diverse datasets. Ensuring the consistency of CSA inference via segmentation analysis remains an open theoretical question that we intend to address in future work. Improvements in functional inference could have an immediate impact on modern single-particle tracking investigations in intracellular biophysics.

The preliminary application to EB1-GFP microtubule plus-end trajectories in neurons (Section~\ref{sec:microtubule}) highlights a third direction for future work: extending CPLASS 
to non-linear piecewise paths suited to the curved and branched 
geometry of neuronal dendrites. In its current piecewise-linear 
form, CPLASS already produces biologically plausible and 
reproducible segmentations of microtubule polymerization 
dynamics, with inferred speeds consistent with known growth 
rates in neuronal dendrites. A generalization to non-linear paths would enable direct association of local
polymerization speed with cellular geometry, connecting our statistical
framework with emerging theoretical models of microtubule organization
\citep{Nelson2024MinimalMechanisms, Scanlon2025NucleationFeedback}.
\section*{Use of AI tools declaration}
The authors used ChatGPT and Claude to assist with code debugging and optimization. All AI-assisted outputs were reviewed and verified by the authors, who take full responsibility for the content of this article.

\section*{Data and code}\label{data-availability-statement}
\vspace{-0.5cm}
The \texttt{cplass} R package implementing the CPLASS algorithm is publicly
available at \url{https://github.com/linhdo154/cplass}. The repository includes
source code, documentation, tests, illustrative examples, and installation
instructions. A companion documentation website is available at
\url{https://linhdo154.github.io/cplass/}.

\section*{Acknowledgments}

This research was supported by the NSF-Simons Southeast Center for Mathematics and Biology (SCMB) through the grants National Science Foundation DMS1764406 and Simons Foundation/SFARI 594594. This work was also supported by National Institutes of Health grant R01NS121245. We thank Christine K.~Payne, William O.~Hancock, and Melissa M.~Rolls for many substantive discussions and sharing insights and experimental data results from their labs.

 \bibliography{bib}
 \appendix
\section*{Supplemental information}

\section{Additional figures}\label{Apd:addition_figures}

This section provides additional CPLASS output for simulated and experimental trajectories. For visualization, segments with inferred speed below $0.1\mu$m/s are shaded pink and segments with inferred speed above $0.1\mu$m/s are shaded green. This threshold is used only for visualization and not by the CPLASS segmentation algorithm.


\subsection{Examples of CPLASS performance for experimental observations}

\begin{figure}[htb!]
      \centering
      \subcaptionbox*{(A) \par} {\includegraphics[width = 0.49\textwidth]{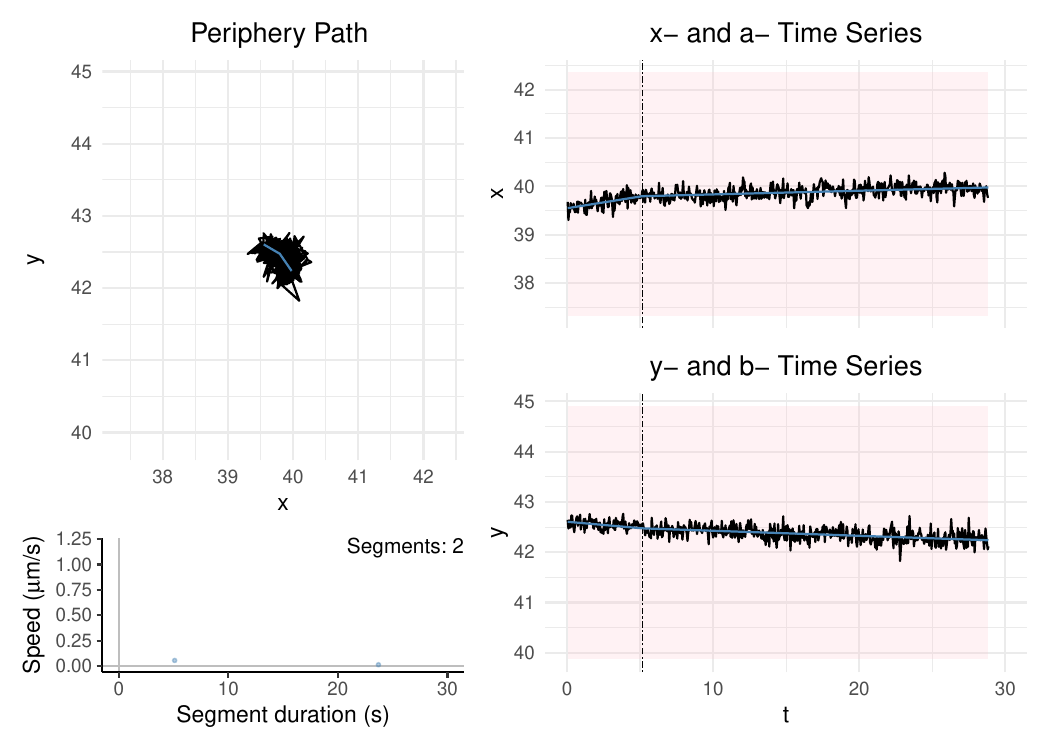}}
      \subcaptionbox*{(B)\par}{\includegraphics[width = 0.49\textwidth]{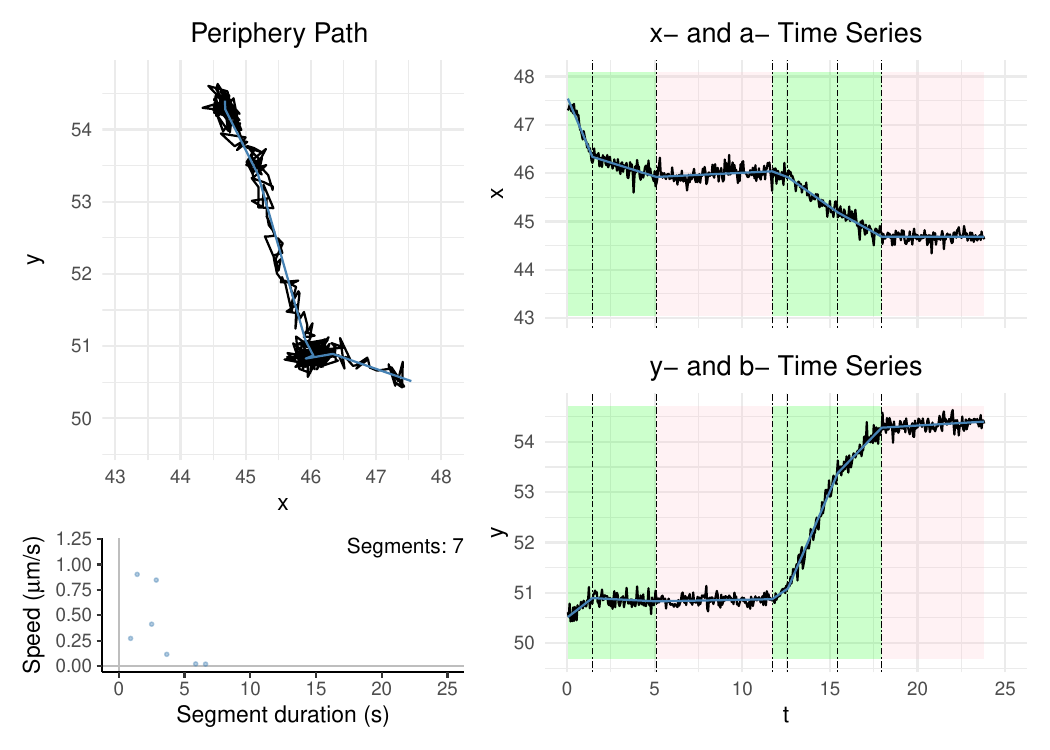}}
      \subcaptionbox*
      {(C) \par} {\includegraphics[width = 0.49\textwidth]{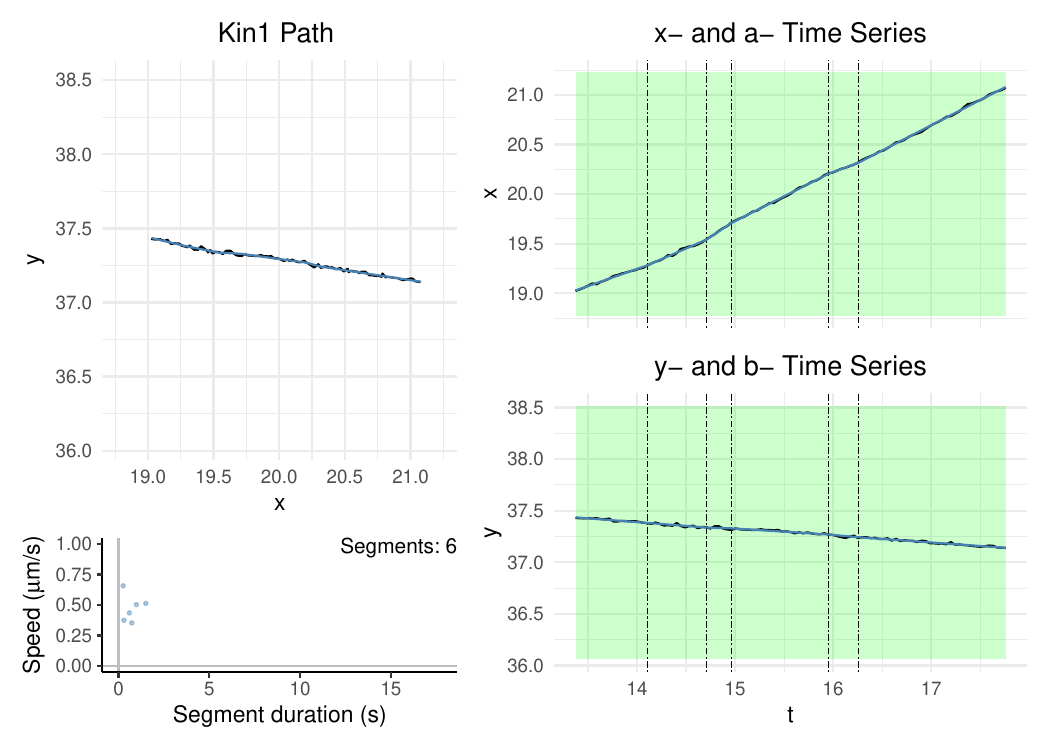}}
      \subcaptionbox*{(D)\par}{\includegraphics[width = 0.49\textwidth]{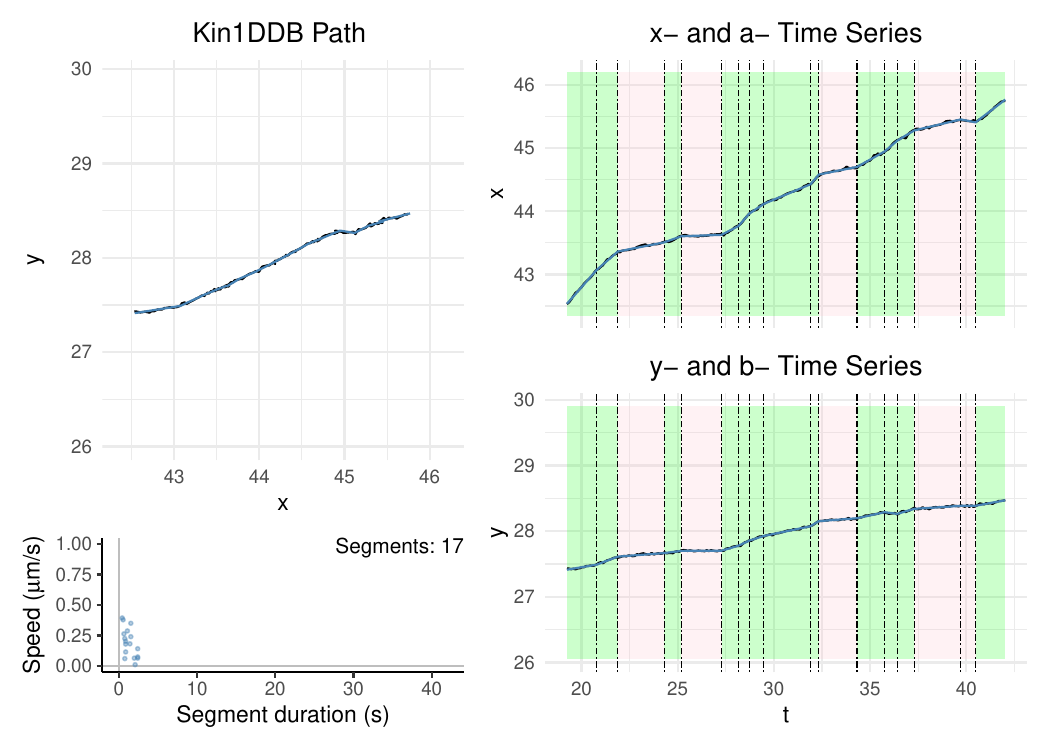}}
      \caption{\textit{CPLASS performance for intracellular transport trajectories} \textbf{Panel (A) and (B)} - \emph{in vivo} lysosomal transport - examples of a non-active path (Panel A) and an active path (Panel B). Panel (C) and (D) - \emph{in vitro} quantum dot transport - are examples of an active path with a kin1 motor (Panel C) and a path moving and pausing while the two motors kin1 and DDB are present at the same time.} \label{fig:real_data_paths}
\end{figure}

\begin{figure}[htb!]
    \centering
    \includegraphics[width=1\linewidth]{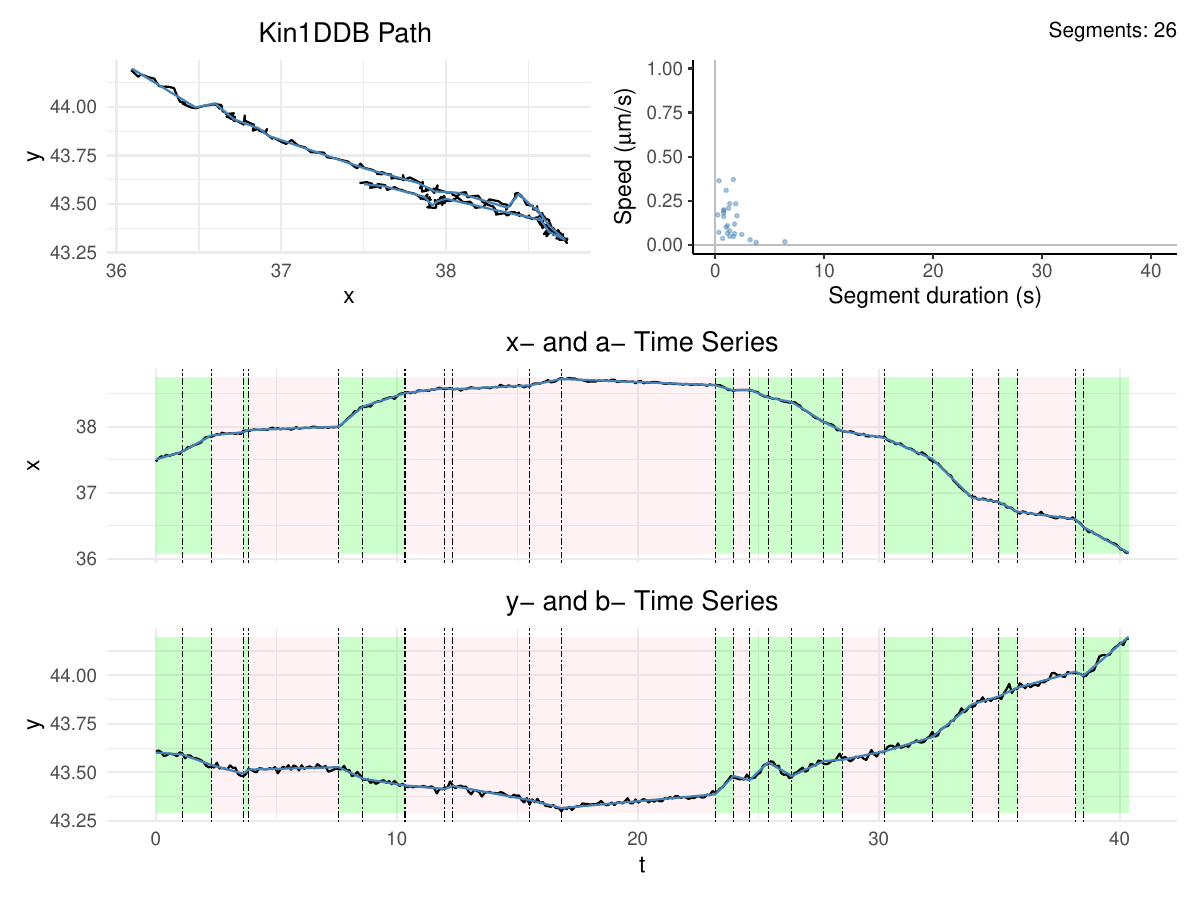}
    \caption{CPLASS analysis of a kinesin-1/DDB trajectory exhibiting bidirectional transport. The green and pink shaded regions indicate segments that are estimated to have a speed of over or under 100 nm/s respectively.}
    \label{fig:kin1DDB_path13}
\end{figure}

In this section, we provided some real paths and output from CPLASS, Figure~\ref{fig:real_data_paths} shows four paths coming from different datasets, the first two panel (A) and (B) are paths from the lysosomal transport dataset in the periphery region of the cell, the last two panel (C) and (D) are paths from the quantum dot transport datasets under two circumstances - one with a single Kinesin-1 motor (panel (C)) and one when the two motors Kin1 and DDB present at the same time. Figure~\ref{fig:kin1DDB_path13} is another path in the group Kin1DDB. We can observe bidirectional, multi-state transport, possibly due to tug-of-war dynamics between the motors. For the Kin1 path, the motor-cargo complex appears to be always in active transport; the detected changes in this case are when it changes the speed or direction during the movement. 

\subsection{BCP and CPLASS on detecting short segments}\label{apd:bcp_cplass}

\begin{figure}[htb!]
    \centering
    \includegraphics[width=1\linewidth]{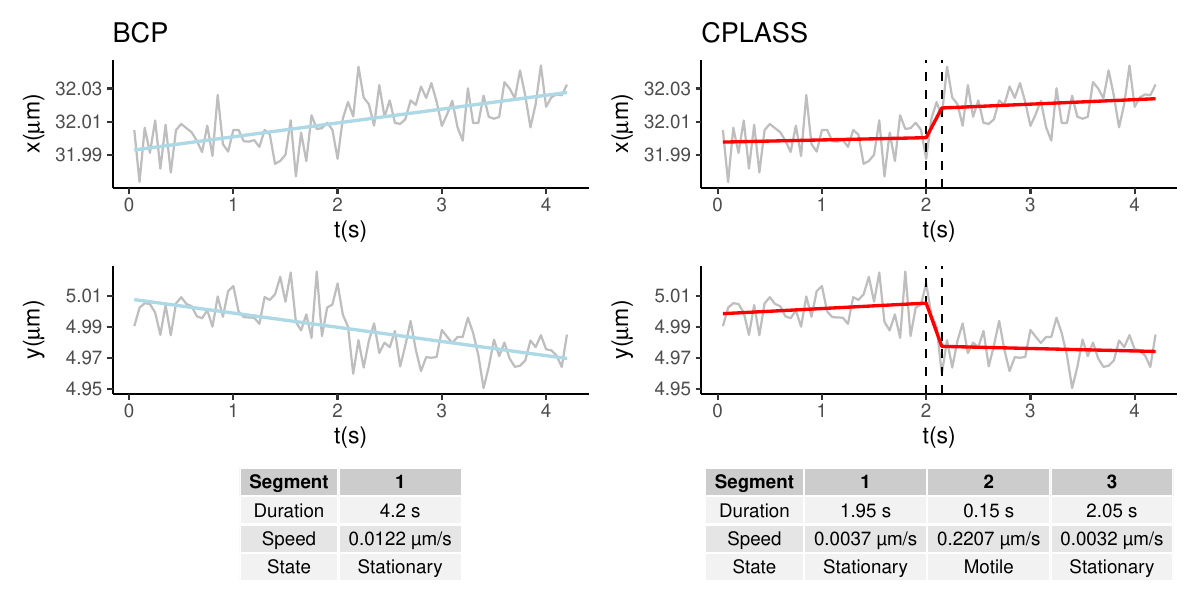}
    \caption{\textit{Comparison of BCP and CPLASS for detection of a short active-transport segment}. The trajectory is one of the simulated paths analyzed in the main manuscript, observed at 20 Hz with $n=84$ and $\sigma=0.01$. Two true changepoints define three segments with durations ($2,0.2,2)s$ and speeds $(0,0.17,0)\mu$m/s. BCP fails to detect either changepoint, whereas CPLASS identifies both changepoints and estimates the duration and speed of the middle segment close to their true values.}
    \label{fig:BCP_CPLASS_short_segment}
\end{figure}

In this section, we provide an example where BCP struggles with detecting short segments as discussed in the main manuscript (Section 3.1) and Figure 7 (the corresponding figure in the main manuscript). With this issue, the motile segment was missed, and the whole path was labeled stationary. Meanwhile, CPLASS detected the short fast segment and labeled the middle segment active (see Figure~\ref{fig:BCP_CPLASS_short_segment}).

\section{Simulation parameters used in the main manuscript} \label{apd: table}

Table~\ref{tab:parameters}
summarizes the parameter values used to generate 250 synthetic trajectories observed at 25 Hz. Additional details regarding the simulation model are provided by \citet{cook2024considering}. Here, we summarize the model and data generation.

The simulated trajectories in the referenced study are generated using a two-state stochastic model representing cargo motion that alternates between \textit{stationary} and \textit{motile} states. The model is parameterized by transition probabilities between these states, motile speed distributions, and additive noise terms.

During simulation, the cargo switches between states at each discrete time step $\Delta$ according to transition probabilities:
\begin{itemize}
    \item $p$: probability of switching from stationary $\rightarrow$ motile,
    \item $q$: probability of switching from motile $\rightarrow$ stationary.
\end{itemize}
When in the motile state, the cargo speed is sampled from a Gamma distribution with shape $\alpha$ and rate $\beta$. The direction of motion may reverse or persist with probabilities $P_{\mathrm{reverse}}$ and $P_{\mathrm{continue}}$, respectively. A Gaussian noise term with standard deviation $\sigma_{\mathrm{cargo}}$ is added to model cargo localization uncertainty. The total number of simulated points is $n$, and the final simulation time is $t_f = n \Delta$.

The authors define three parameter groups (``Base'', ``Contrast'', and ``Mimic''). The ``Base'' group serves as the reference setting used to simulate synthetic lysosome trajectories at multiple frame rates. The corresponding parameters are summarized in Table~\ref{tab:parameters} below.

{\fontsize{10}{9}\selectfont
\begin{table}[ht!]
\centering
\begin{tabular}{lcl}
\toprule
\textbf{Parameter} & \textbf{Base}  & \textbf{Description} \\
\midrule
$n$ & 200 & Number of Observations \\
$p$ & 1 & Probability \textit{Stationary} to \textit{Motile} \\
$q$ & 0.5  & Probability \textit{Motile} to \textit{Stationary} \\
$\alpha$ & 8 & Speed Shape Parameter \\
$\beta$ & 0.02 & Speed Rate Parameter \\
$\bar{D}$ & 300~nm  & Average Distance Traveled \\
$\sigma$ & 5~s  & Average \textit{Stationary} Duration \\
$P_{\mathrm{reverse}}$ & 0.3 & Probability of reversal \\
$P_{\mathrm{continue}}$ & 0.3 & Probability of same direction \\
$\sigma_{\mathrm{cargo}}$ & 0.1 & Noise magnitude of Cargo \\
$\Delta$ & 0.04~s & Time Step \\
$t_f$ & $n\Delta$  & Final simulation time \\
\bottomrule
\end{tabular}
\caption{Parameters used to model simulated lysosome trajectories. "Base" denotes parameters for simulated lysosome trajectories at any frame rate. In this paper, we use the simulated data at 25Hz, i.e., $\Delta = 0.04$. \label{tab:parameters}}
\end{table}
}

To generate data, the simulation proceeds as follows:
\begin{enumerate}
    \item Initialize the trajectory in either state.
    \item For each time step $\Delta$, update the state based on $p$ and $q$.
    \item If motile, draw a speed from $\mathrm{Gamma}(\alpha, \beta)$ and move the particle accordingly, allowing reversals or continuations based on $P_{\mathrm{reverse}}$ and $P_{\mathrm{continue}}$.
    \item Add Gaussian noise with magnitude $\sigma_{\mathrm{cargo}}$.
    \item Because the segments are not exponentially distributed and the process is non-Markovian, a burn-in period is required and the process is repeated until a total time $t_f$ is reached.
\end{enumerate}

This process yields a set of synthetic lysosome trajectories that serve as ground truth for evaluating segmentation and motion-state inference algorithms.

\section{Checking the detailed balance condition in the MH algorithm}\label{subsec:detailed_balance}

Given the proposal function with four proposal types in the main manuscript (MH search algorithm section), to prove detailed balance for the MH algorithm, we need to show that the transition kernel satisfies:
\begin{equation}\label{eq: detailed_balance}
    \pi(r^{\cur})q_r(r^{\prop}|r^{\cur})\alpha(r^{\prop}|r^{\cur}) = \pi(r^{\prop})q_r(r^{\cur}|r^{\prop})\alpha(r^{\cur}|r^{\prop}),
\end{equation}
where 
\begin{itemize}
    \item $\pi(r) = \exp(\Phi(r))$ is the target posterior distribution of changepoints.
    \item $q_r(r^{\prop}|r^{\cur})$ is the overall proposal function, combining four different proposal types with predefined probabilities, and
    \item $\alpha(r^{\prop}|r^{\cur})$ is the MH acceptance probability:
    $$\alpha\left(r^{\prop}|r^{\cur}\right) = \min \left(1, \dfrac{\pi(r^{\prop})q_r\left(r^{\cur}|r^{\prop}\right)}{\pi(r^{\cur})q_r\left(r^{\prop}|r^{\cur}\right)}\right).$$
\end{itemize}

In order to verify the detailed balance condition in the MH algorithm, we need to analyze whether the proposal function $q_r(r^{\prop}|r^{\cur})$ satisfies symmetry, meaning that the probability of proposing $r^{\prop}$ given $r^{\cur}$ is equal to the probability of proposing $r^{\cur}$ given $r^{\prop}$, or if any asymmetry exists, it is properly accounted for in the acceptance probability.

We analyze the acceptance probability of each type of proposal in detail:

\paragraph{Type 1: $q_{\mathrm{new}}$ (Independent changepoint vector proposal)}
In this proposal, we generate a completely new set of changepoints independently of the current state $r^{\cur}$. The new changepoints are generated from a Bernoulli process with probability $1-\exp(-\lambda\Delta)$. 

We have
{\fontsize{12}{8}\selectfont\begin{align*}
    q_{\new}(r^{\prop}|r^{\cur}) & = \left(\dfrac{1-e^{-\lambda\Delta}}{e^{-\lambda \Delta}}\right)^{|r^{\prop}|} \times (e^{-\lambda \Delta})^{n-1},\\
    q_{\new}(r^{\cur}|r^{\prop}) &= \left(\dfrac{1-e^{-\lambda\Delta}}{e^{-\lambda \Delta}}\right)^{|r^{\cur}|} \times (e^{-\lambda \Delta})^{n-1}.
\end{align*}}
Therefore, 
\begin{align*}
\alpha(r^{\cur}|r^{\prop}) = \min\left(1, \dfrac{\pi(r^{\prop})}{\pi(r^{\cur})}\times\left(\dfrac{e^{-\lambda\Delta}}{1-e^{-\lambda\Delta}}\right)^{|r^{\prop}|-|r^{\cur}|}\right),
\end{align*}

\paragraph{Type 2: $q_{\bd}$ (Birth/death proposal)}
This proposal adds or removes a single changepoints at random.
We have that if $q_{\bd}(r^{\prop}|r^{\cur})=\dfrac{1}{2|r^{\cur}|}$ then $q_{\bd}(r^{\cur}|r^{\prop}) = \dfrac{1}{2(n-|r^{\prop}|-1)}$ and vice versa. The acceptance rate is then
\begin{align*}
    \alpha(r^{\cur}|r^{\prop}) &= \min\left(1, \dfrac{\pi(r^{\prop})}{\pi(r^{\cur})}\times\dfrac{|r^{\cur}|}{n-|r^{\prop}|-1}\right) \text{ or }\\ \alpha(r^{\cur}|r^{\prop}) &= \min\left(1, \dfrac{\pi(r^{\prop})}{\pi(r^{\cur})}\times\dfrac{n-|r^{\cur}|-1}{|r^{\prop}|}\right),
\end{align*}
respective.
\paragraph{Type 3: $q_{\bd_2}$ (Segment insertion/deletion proposal)} For this type of proposal, if $q_{\bd_2}(r^{\prop}|r^{\cur})= \dfrac{1}{|r^{\cur}|
      }\one_{\{|r^{\cur}| \ge 2\}
      } $ then  
      $$q_{\bd_2}(r^{\cur}|r^{\prop}) = \dfrac{1}{2}\sum_{j=1}^{|r^{\prop}|+1} \dfrac{(\dd_j-1)(\dd_j-2)}{(n-|r^{\prop}|-1)(n-|r^{\prop}|-2)}$$ and vice versa. The acceptance rate is then 
{\fontsize{10}{10}\selectfont
\begin{align*}
    \alpha(r^{\cur}|r^{\prop}) &= \min\left(1, \dfrac{\pi(r^{\prop})}{\pi(r^{\cur})} \times 
    |r^{\cur}|\times\one_{\{|r^{\cur}|\ge 2\}}\times \dfrac{1}{2}\sum_{j=1}^{|r^{\prop}|+1} \dfrac{(\dd_j-1)(\dd_j-2)}{(n-|r^{\prop}|-1)(n-|r^{\prop}|-2)}\right)\\
    &\text{or}\\
    \alpha(r^{\cur}|r^{\prop}) &= \min\left(1, \dfrac{\pi(r^{\prop})}{\pi(r^{\cur})} \times \dfrac{\dfrac{1}{|r^{\prop}|
      }\one_{\{|r^{\prop}| \ge 2\}
      } }{\dfrac{1}{2}\sum_{j=1}^{|r^{\cur}|+1} \dfrac{(\dd_j-1)(\dd_j-2)}{(n-|r^{\cur}|-1)(n-|r^{\cur}|-2)}}\right), \text{ respectively.}
\end{align*}}

\paragraph{Type 4: $q_{\shift}$ (Shift proposal)} This proposal shifts the position of one of the changepoints in the current list of changepoints. The proposal is symmetric as defined. Therefore, the acceptance rate in this case is:
$\alpha(r^{\cur}|r^{\prop}) = \min\left(1, \dfrac{\pi(r^{\prop})}{\pi(r^{\cur})}\right)$.

Given all the acceptance rates for each type of proposal, there are two cases which are

\paragraph{Case 1: $\pi(r^{\prop})q_{\new}(r^{\cur}|r^{\prop}) \ge \pi(r^{\cur})q_{\new}(r^{\prop}|r^{\cur})$.} We then have  
\begin{align*}
    \alpha(r^{\prop}|r^{\cur}) = 1; \quad \alpha(r^{\cur}|r^{\prop}) = \dfrac{\pi(r^{\cur})q_{\new}\left(r^{\prop}|r^{\cur}\right)}{\pi(r^{\prop})q_{\new}\left(r^{\cur}|r^{\prop}\right)}. 
\end{align*}

Plugging into the two sides of Equation~\eqref{eq: detailed_balance}, we get that both sides of this equation are equal to $\pi(r^{\cur})q_{\new}(r^{\prop}|r^{\cur})$.

\paragraph{Case 2: $\pi(r^{\prop})q_{\new}(r^{\cur}|r^{\prop}) <\pi(r^{\cur})q_{\new}(r^{\prop}|r^{\cur})$.} We then have 
\begin{align*}
    \alpha(r^{\cur}|r^{\prop}) = 1; \quad \alpha(r^{\prop}|r^{\cur}) = \dfrac{\pi(r^{\prop})q_{\new}\left(r^{\cur}|r^{\prop}\right)}{\pi(r^{\cur})q_{\new}\left(r^{\prop}|r^{\cur}\right)}. 
\end{align*}

Plugging into the two sides of Equation~\eqref{eq: detailed_balance}, we get that both sides of this equation are equal to $\pi(r^{\prop})q_{\new}(r^{\cur}|r^{\prop})$.

We finish checking the detailed balance condition.

\section{Computational performance}

To characterize the computational performance of CPLASS, we benchmarked the current R package implementation using simulated trajectories. Each benchmark was repeated five times, and median elapsed times
together with interquartile ranges are reported.

\begin{figure}[!h]
    \centering
    \includegraphics[width=1\linewidth]{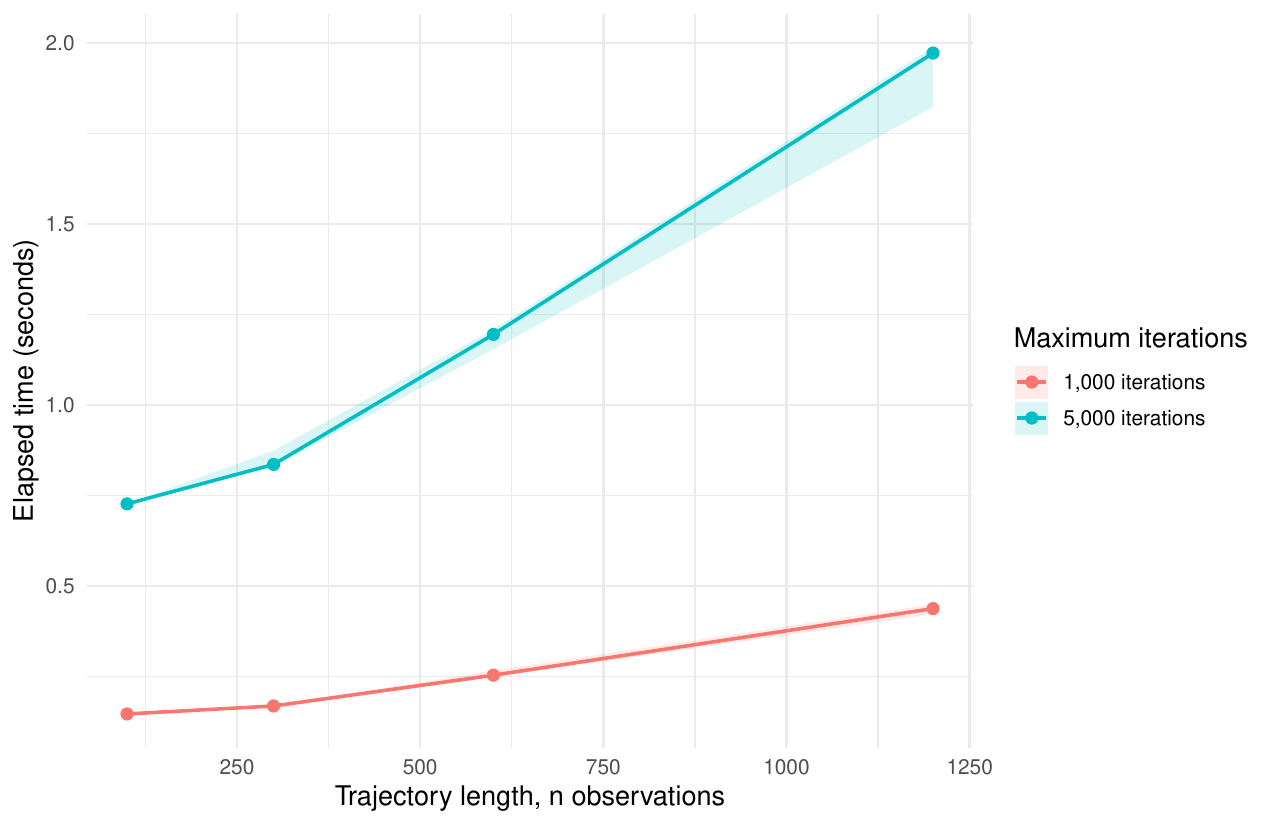}
    \caption{
Single-chain runtime of CPLASS as a function of trajectory length.
Synthetic trajectories containing 100--1200 observations were analyzed using
either 1,000 or 5,000 maximum MCMC iterations. Points represent the median
elapsed time across five independent runs, and shaded regions denote the
interquartile range. Runtime increased gradually with both trajectory length
and the number of MCMC iterations while remaining below two seconds for
trajectories containing up to 1,200 observations using 5,000 iterations.
}\label{fig:single_chain}
\end{figure}

For single-chain analyses, runtime increased gradually with trajectory length
and with the maximum number of MCMC iterations (Supplementary Figure~\ref{fig:single_chain}). For
synthetic trajectories containing between 100 and 1200 observations, a
5,000-iteration CPLASS run required a median of only 0.73--1.97 seconds,
whereas a 1,000-iteration run required 0.15--0.44 seconds.

\begin{figure}[!h]
    \centering
    \includegraphics[width=1\linewidth]{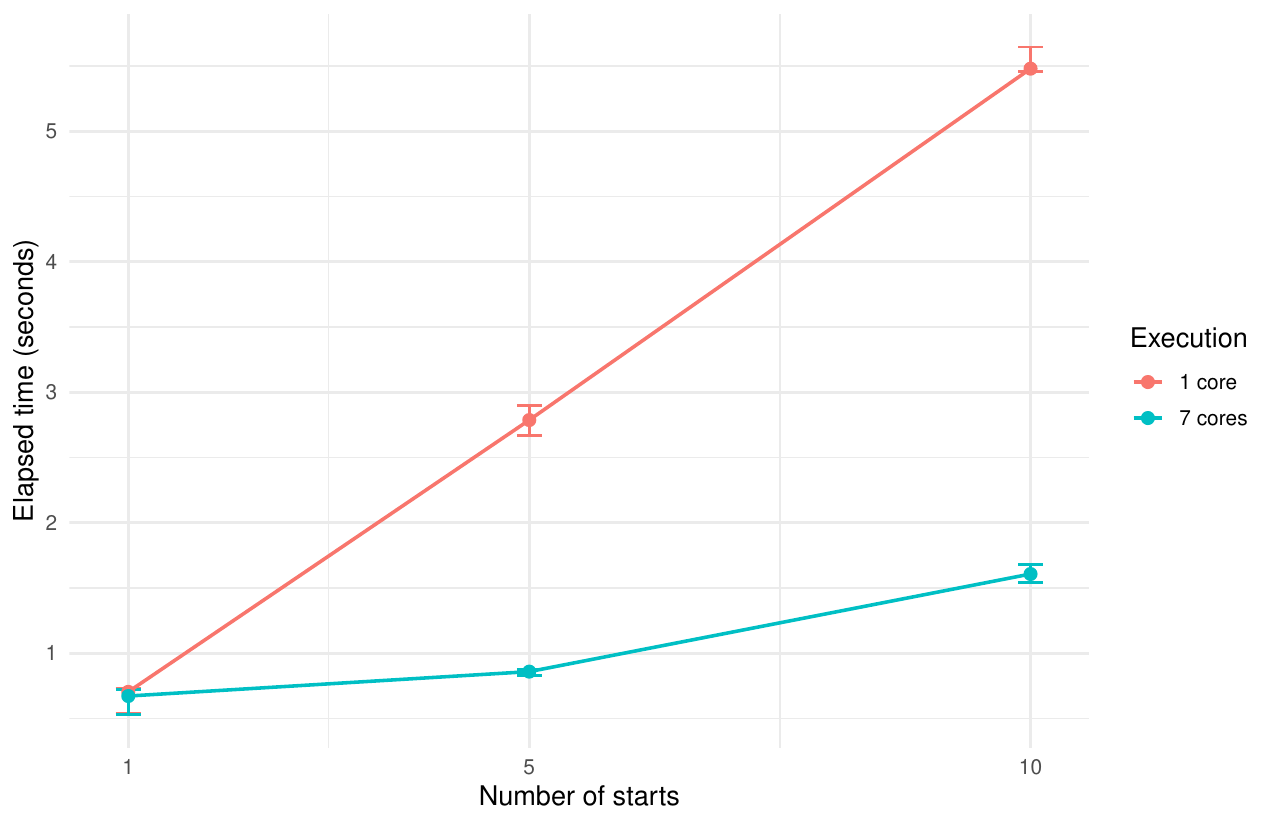}
    \caption{
Runtime of the CPLASS multistart algorithm using a synthetic trajectory with 600 observations and a maximum of 3,000 MCMC iterations per chain. Results are shown for serial execution (1 CPU core) and parallel execution (7 CPU cores). Points represent the median elapsed time across five repeated runs, and error bars denote the interquartile range. Parallel execution substantially reduced total runtime, providing approximately 3.2-fold and 3.4-fold speedups for five and ten independent starts, respectively.
}
    \label{fig:runtime_multistart}
\end{figure}

We next evaluated the multistart implementation using synthetic trajectories of 600 observations with a maximum of 3,000 iterations per chain (Supplementary Figure~\ref{fig:runtime_multistart}). Parallel execution substantially reduced computation time. For example, ten independent starts required a median of 5.48 seconds using a single CPU core but only 1.61 seconds using seven CPU cores, corresponding to an approximately 3.4-fold speedup.

\newpage

\end{document}